\documentclass[times,twocolumn]{elsarticle}

\usepackage{medima}
\usepackage{framed,multirow}

\usepackage{amssymb}
\usepackage{latexsym}

\usepackage{dirtree}
\usepackage{url}
\usepackage{epsfig}
\usepackage{graphicx}
\usepackage{amsmath}
\usepackage{amssymb}
\usepackage{array}
\usepackage{color}
\usepackage[ruled]{algorithm}
\usepackage{algorithmic}
\usepackage{setspace}
\usepackage{fullpage}
\usepackage{mathtools}
\usepackage{float}
\usepackage{hyperref}
\usepackage{rotating}

\definecolor{newcolor}{rgb}{.8,.349,.1}

\journal{Medical Image Analysis}

\begin{document}

\verso{Given-name Surname \textit{et~al.}}

\begin{frontmatter}

\title{Deep reinforcement learning in medical imaging: A literature review}

\author[1]{S. Kevin \snm{Zhou}\corref{cor1}}
\cortext[cor1]{Corresponding author.}
\author[2]{Hoang Ngan \snm{Le}\corref{cor1}}
\author[2]{Khoa \snm{Luu}}
\author[3]{Hien \snm{V. Nguyen}}
\author[4]{Nicholas \snm{Ayache}}

\address[1]{MIRACLE, Institute of Computing Technology, Chinese Academy of Sciences}
\address[2]{CSCE Department, University of Arkansas}
\address[3]{ECE Department, University of Houston}
\address[4]{INRIA, Sophia Antipolis-Mediterranean Centre}

\begin{abstract}
Deep reinforcement learning (DRL) augments the reinforcement learning framework, which learns a sequence of actions that maximizes the expected reward, with the representative power of deep neural networks. Recent works have demonstrated the great potential of DRL in medicine and healthcare.
This paper presents a literature review of DRL in medical imaging. We start with a comprehensive tutorial of DRL, including the latest model-free and model-based algorithms. We then cover existing DRL applications for medical imaging, which are roughly divided into three main categories: (i) parametric medical image analysis tasks including landmark detection, object/lesion detection, registration, and view plane localization; (ii) solving optimization tasks including hyperparameter tuning, selecting augmentation strategies, and neural architecture search; and (iii) miscellaneous applications including surgical gesture segmentation, personalized mobile health intervention, and computational model personalization. The paper concludes with discussions of future perspectives.
\end{abstract}


\end{frontmatter}

\section{Introduction}
\label{sec:intro}

Reinforcement learning is a framework for learning a sequence of actions that maximizes the expected reward \cite{sutton2018reinforcement,li2017deep}. Deep reinforcement learning (DRL) is the result of marrying deep learning with reinforcement learning \cite{mnih2013playing}. DRL allows reinforcement learning to scale up to previously intractable problems. Deep learning and reinforcement learning were selected by MIT Technology Review as one of 10 Breakthrough Technologies\footnote{https://www.technologyreview.com/10-breakthrough-technologies/} in 2013  and 2017, respectively. The combination of these two powerful technologies currently constitutes one of the state-of-the-art frameworks in artificial intelligence.

Recent years have witnessed rapid progress in DRL, resulting in significant performance improvement in many areas, including games \cite{mnih2013playing}, robotics \cite{finn2016guided}, natural language processing \cite{luketina2019survey}, and computer vision \cite{bernstein2018reinforcement}. Unlike supervised learning, DRL framework can deal with sequential decisions, and learn with highly delayed supervised information (e.g., success or failure of the decision is available only after multiple time steps). DRL can also deal with non-differentiable metrics. For example, one can use DRL to search for an optimal deep network architecture \cite{zoph2016neural} or parameter settings to maximize the classification accuracy, which is clearly non-differentiable with respect to the number of layers or the choice of non-linear rectifier functions. Another use of DRL is in finding efficient search sequence for speeding up detection, or optimal transformation sequence for improving registration accuracy. DRL can also mitigate the issue of high memory consumption in processing high-dimensional medical images. For example, a DRL-based object detection can focus on a small image region at a time, which incurs a lower memory footprint, then decide next regions to process.

Despite its successes, application of this DRL technology to medical imaging remains to be fully explored \cite{zhou2020review}. This is partly due to the lack of a systematic understanding of the DRL's strengths or weaknesses when applying to medical data. To this end, we organized a MICCAI 2018 tutorial \footnote{The tutorial is available online at \href{https://www.hvnguyen.com/deepreinforcementlearning}{https://www.hvnguyen.com/deepreinforcementlearning}}, with its
goal of bridging the gap by providing a comprehensive introduction to deep reinforcement learning methods in terms of theories, practice, and future directions. The tutorial contained multiple presentations from active researchers in DRL, covering state-of-the-art and explaining in-depth how DRL was applied in a selected set of topics such as neural architecture search \cite{zoph2016neural}, detection \cite{ghesu2016artificial}, segmentation \cite{sahba2006reinforcement}, and controlling of surgical robots \cite{liu2018deep}. This tutorial forms the basis of the paper. However, in this paper we go much beyond the tutorial and expand it with many state-of-the-art contents. 

Our goal is to provide our readers good knowledge about of the principle of DRL and a thorough coverage of the latest examples of how DRL is used for solving medical imaging tasks. We structure the rest of paper as follows: (i) introduction to deep reinforcement learning with its generation framework and latest learning strategies; (ii) how to use DRL for solving medical image analysis tasks, which is the main part that covers the literature review; (iii) fundamental challenges and future potential of DRL in medical domains. 

\section{Basics of Reinforcement Learning}
\label{sec:rl}
This section serves as a brief introduction to the theoretical models and techniques in RL. In order to provide a quick overview of what constitutes the main components of RL methods, some fundamental concepts and major theoretical problems are also clarified.

RL is a kind of machine learning methods where agents learn the optimal policy by trial and error. Inspired by behavioral psychology, RL was proposed for sequential decision-making tasks which are potential to many applications such as robotics, healthcare, smart grids, finance, self-driving cars, etc. Similar to a biological agent, an artificial agent collects experiences by interacting with its environment. Such experience is then gathered to optimize some objectives given in the form of cumulative rewards.

\begin{figure}[htbp]
	\centering \includegraphics[width=\columnwidth]{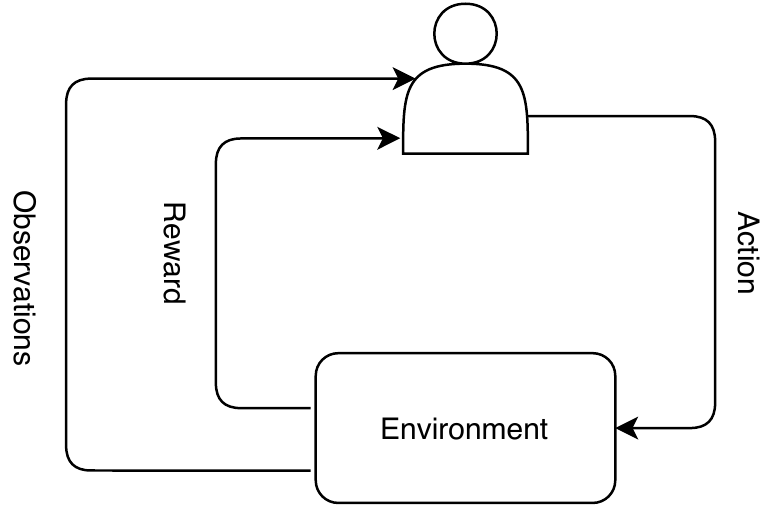}
	\caption{An illustration of agent-environment interaction in RL.}
	\label{fig:RL}
\end{figure}

Here we focus on how the RL problem can be formalized as an agent that is able to make decisions in an environment to optimize some objectives.
Key aspects of RL include: (i) Addressing the sequential decision making; (ii) There is no supervisor, only a reward presented as scalar number; (iii) Feedback is highly delayed. The interaction between agent and environment is illustrated in Fig. \ref{fig:RL}.

\subsection{Markov Decision Process}
The standard theory of RL is defined by a Markov Decision Process (MDP). A discrete time stochastic process whose conditional probability distribution of the future states only depends upon the present state is called Markov Process (also known as Markov chain). By introducing the concept of reward and action into Markov Process, Markov Process can be extended to MDP. In MDP, the immediate reward obtained at the future depends not only on the current state but also on the action that leads to the future state. An MDP is typically defined by  five elements as follow:
\begin{itemize}
    \item $S$: a set of \textit{state}/observation space of an environment. $s_0$ is a starting state.
    \item $\mathcal{A}$: set of \textit{actions} the agent can choose from.
    \item $T$: a \textit{transition probability} function $T(s_{t+1}| s_t, a_t)$, specifying the probability that the environment will transition to state $s_{t+1}\in S$ if the agent takes action $a\in \mathcal{A}$ in state $s \in S$.
    \item $R$:  a \textit{reward} function where $r_{t+1} = R(s_t, s_{t+1})$ is a reward received for taking action $a_t$ at state $s_t$ and transfer to the next state $s_{t+1}$.
    \item $\gamma$: discount factor.
\end{itemize}

Considering MDP($S$, $\mathcal{A}$, $\gamma$, $T$, $R$), the agent chooses an action $a_t$ according to the policy $\pi(a_t|s_t)$ at state $s_t$. Notably, the agent's algorithm for choosing an action $a$ given its current state $s$, which in general can be viewed as distribution $\pi(a | s)$, is called a \textit{policy} (strategy). The environment receives the action, produces a reward $r_{t+1}$ and transfers to the next state $s_{t+1}$ according to the transition probability $T(s_{t+1}|s_t, a_t)$. The process continues until the agent reaches a terminal state or a maximum time step. In RL framework, the 4-tuple $(s_t, a_t, r_{t+1}, s_{t+1})$ is called \textit{transition}. Several sequential transitions are usually referred to as roll-out. A full sequence $(s_0, a_0, r_1, s_1, a_1, r_2, ... )$ is called a \textit{trajectory}. Theoretically,  a trajectory goes to infinity, but the episodic property holds in most practical cases. One trajectory of some finite length $\tau$, is called an \textit{episode}. For given MDP and policy $\pi$, the probability of observing $(s_0, a_0, r_1, s_1, a_1, r_2, ... )$  is called \textit{trajectory distribution} and is denoted as:
\begin{equation}
    \mathcal{T}_{\pi} = \prod_{t}{\pi(a_t|s_t)T(s_{t+1}|s_t, a_t)}.
    \label{eq:T0}
\end{equation}

\noindent The objective of RL is to find the \textit{optimal policy} $\pi^*$ for the agent that maximizes the cumulative reward, called \textit{return}. For every episode, return is defined as the weighted sum of immediate rewards:
\begin{equation}
 \mathcal{R} = \sum_{t=0}^{\tau-1}{\gamma^t r_{t+1}}.
 \label{eq:R}
\end{equation}
\noindent Because the policy induces a trajectory distribution, the \textit{expected reward} maximization can be written as:
\begin{equation}
\mathbb{E}_{{\mathcal{T}_\phi}} \sum_{t=0}^{\tau-1}{r_{t+1}} \rightarrow \max_{\pi}.
\end{equation}

\noindent Thus, given MDP and policy $\pi$, the \textit{discounted expected reward} is defined:
\begin{equation}
    \mathcal{G}(\pi) = \mathbb{E}_{{\mathcal{T}_\phi}}\sum_{t=0}^{\tau-1}\gamma^{t}{r_{t+1}}.
\end{equation}
The goal of RL is to find an \textit{optimal policy} $\pi^*$, which maximizes the discounted expected reward, i.e. $\mathcal{G}(\pi) \rightarrow \max_{\pi}$

\subsection{Value functions}
In order to estimate how good it is for an agent to utilize policy $\pi$ to visit state $s$, a value function is introduced. The value is the mathematical expectation of return and value approximation is obtained by Bellman expectation equation as follows:
\begin{equation}
    V^\pi(s_t) = \mathbb{E}[r_{t+1} + \gamma V^\pi(s_{t+1})]. \label{eq:value}
\end{equation}
$V^\pi(s_t)$ is also known as state-value function, and the expectation term can be expanded as a product of policy, transition probability, and return as follows:
\begin{align}
    & V^\pi(s_t)  = \nonumber\\ & \sum_{a_t\in\mathcal{A}}{\pi(a_t|s_t)}\sum_{s_{t+1}\in
    S}{T(s_{t+1} | s_t, a_t)[R(s_t, s_{t+1}) + \gamma V^\pi(s_{t+1})]}.
\end{align}
This equation is called Bellman equation. When the agent always selects the action according to the optimal policy $\pi^*$ that maximizes the value, Bellman equation can be expressed as following:
\begin{align}
V^*(s_t) &=& \max_{a_t}\sum_{s_{t+1}\in S}{T(s_{t+1} | s_t, a_t)[R(s_t, s_{t+1}) + \gamma V^*(s_{t+1})]} \nonumber\\
&\overset{\Delta}{=}& \max_{a_t}Q^*(s_t, a_t). \label{eq:bestvalue}
\end{align}

However, obtaining optimal value function $ V^*$ does not provide enough information to reconstruct some optimal policy $\pi^*$ because of the complexity of the real world. Thus, a quality function (Q-function) under policy $\pi$ is introduced as:
\begin{equation}
    Q^\pi(s_t, a_t) = \sum_{s_{t+1}}{T(s_{t+1}|s_t, a_t)[R(s_t, s_{t+1})+\gamma V^\pi(s_{t+1})]}.
\end{equation}

\subsection{Category}
In general, RL can be divided into model-free model-based methods.  Here, ``model" refers to the environment itself that is defined by the two quantities: transition probability function $T(s_{t+1} | s_t, a_t)$ and reward function $R(s_t, s_{t+1})$.

\subsubsection{Model-based methods}
Such methods exploit learned or given world dynamics, \textit{i.e.}, $T(s_{t+1} | s_t, a_t)$, $R(s_t, s_{t+1})$. There are four main model-based techniques as follows:
\begin{itemize}
    \item \underline{Value function.} The objective of value function methods is to obtain the best policy by maximizing the value functions in each state. A value function of a RL problem can be defined as in Eq. (\ref{eq:value}) and the optimal state-value function is given in Eq. (\ref{eq:bestvalue}), which are known as Bellman equations. Some common approaches in this group are differential dynamic programming \cite{DDP_1}, \cite{DDP_2}, temporal difference learning \cite{TDP_1}, policy iteration \cite{PI_1}, and Monte Carlo \cite{MTCL}.

    \item \underline{Transition models.} Transition models decide how to map from a state $s$, taking action $a$ to the next state ($s'$) and it strongly effect the performance of model-based RL algorithms. Depend on whether predicting the future state $s'$ is based on probability distribution of a random variable or not, there are two main approaches in this group: stochastic and deterministic. Some common deterministic methods are decision trees \cite{TM_2} and linear regression \cite{TM_1}. Some common stochastic methods are Gaussian processes \cite{TM_3}, \cite{TM_4}, \cite{TM_5}, expectation-maximization \cite{TM_6}, and dynamic Bayesian networks \cite{TM_2}.
    \item \underline{Policy search.} Policy search approach directly searches for the optimal policy by modifying its parameters whereas the value function methods indirectly find the actions that maximize the value function at each state. There are three approaches in this group:  gradient-based \cite{PS_1}, \cite{PS_2}, information theory \cite{TM_4}, \cite{PS_3}, and sampling based \cite{RP_1}.
    \item \underline{Return functions.} A return function decides how to aggregate rewards or punishments over an episode. It affects both the convergence and the feasibility of the model. There are two main approaches in this group: discounted returns functions \cite{RP_1}, \cite{RP_2}, \cite{RP_3} and averaged returns functions \cite{RP_4}, \cite{RP_5}. Between two approaches, the former is the most popular which represents the uncertainty about future rewards. While small discount factors provide a faster convergence, its solution many not optimal.
\end{itemize}
In practice, transition and reward functions are rarely known and hard to model. The comparative performances among all model-based techniques are reported in \cite{model-basedRL} with over 18 benchmarking environments including noisy ones.

\subsubsection{Model-free methods:} Such methods learn through the experience gained from interactions with the environment, that is, a model-free method tries to estimate the transition probability function and the reward function from the experience to exploit them in acquisition of policy. Policy gradient and value-based algorithms are popularly used in model-free methods.
\begin{itemize}
    \item \underline{The policy gradient methods.} In this approach, RL task is considered as an optimization with stochastic first-order optimization. Policy gradient methods directly optimize the discounted expected reward, \textit{i.e.}, $\mathcal{G}(\pi) \rightarrow \max_{\pi}$ to obtains the optimal policy $\pi^*$ without any additional information about MDP. To do so, approximate estimations of gradient with respect to policy parameters are used. Taking \cite{williams1992simple} as an example, policy gradient parameterizes the policy and updates parameters $\theta$:
    \begin{equation}
        \mathcal{G}^\theta(\pi) = \mathbb{E}_{{\mathcal{T}_\phi}}\sum_{t=0}{log( \pi_\theta(a_t|s_t))\gamma^{t}\mathcal{R}},
        \label{eq:policy_gradient1}
    \end{equation}
    where $\mathcal{R}$ is the total accumulated return defined in Eq. (\ref{eq:R}).

    \item \underline{Value-based methods.} In this approach, the optimal policy $\pi^*$ is implicitly conducted by gaining an approximation of optimal Q-function $Q^*(s, a)$. In value-based methods, agents update the value function to learn suitable policy while policy-based RL agents learn the policy directly. Q-learning is a typical value-based method. The updating rule of Q-learning with a learning rate $\lambda$ is defined as:
    \begin{equation}
        Q(s_t, a_t) = Q(s_t, a_t) + \lambda \delta_t,
    \end{equation}
    where $\delta_t = R(s_t, s_{t+1}) + \gamma \text{arg} \max_a{Q(s_{t+1}, a) - Q(s_{t}, a)}$ is the temporal difference (TD) error.

    \item \underline{Actor-critic} is an improvement of policy gradient with a value-based critic $\Gamma$; thus, Eq. (\ref{eq:policy_gradient1}) is rewritten as:
    \begin{equation}
        \mathcal{G}^\theta(\pi) = \mathbb{E}_{{\mathcal{T}_\phi}}\sum_{t=0}{log( \pi_\theta(a_t|s_t))\gamma^{t}\Gamma_t }.
        \label{eq:actor_crtic}
    \end{equation}
    The critic function $\Gamma$ can be defined as $Q^\pi(s_t, a_t)$ or $Q^\pi(s_t, a_t) - V^\pi_t $ or $R[s_{t-1}, s_{t}] + V^\pi_{t+1} - V^\pi_t$.
\end{itemize}

\begin{figure}
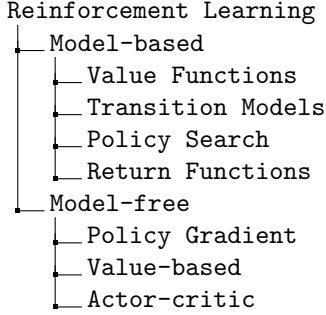

\dirtree{%
.1 Reinforcement Learning.
.2 Model-based.
.3 Value Functions.
.3 Transition Models.
.3 Policy Search.
.3 Return Functions.
.2 Model-free.
.3 Policy Gradient.
.3 Value-based.
.3 Actor-critic.
}
\caption{Summary of RL approaches with model-based and model-free techniques.}
\label{fig:summaryRL}
\end{figure}

Figure \ref{fig:summaryRL} summarizes different RL approaches. The comparison between model-based and model-free methods is given in Table \ref{tb:comparison}.

\begin{table*}[htbp]
\centering
\begin{tabular}{|l|l|l|}
\hline
\textbf{Factors}  & \textbf{Model-based RL}                                                       & \textbf{Model-free RL}                                                              \\ \hline
\begin{tabular}[c]{@{}l@{}}Number of iterations between \\ agent and environment\end{tabular} & Small  & Big  \\ \hline
Convergence & Fast  & Slow \\ \hline
Prior knowledge of transitions & Yes  & No \\ \hline
Flexibility                                        & \begin{tabular}[c]{@{}l@{}}Strongly depends on \\ a learnt model\end{tabular} & \begin{tabular}[c]{@{}l@{}}Adjust based \\ on trials and errors\end{tabular} \\ \hline
\end{tabular}
\caption{Comparison between model-based RL and model-free RL}
\label{tb:comparison}
\end{table*}



\section{Introduction to Deep Reinforcement Learning}
\label{sec:drl}

Thanks to the rich context representation of Deep Learning (DL), DRL was proposed as a combination of RL and DL and has been achieved rapid developments. Under DRL, the aforementioned value and policy can be expressed by a neural network, which allows to deal with a continuous state or action that is hard for a table representation. Similar to RL, DRL can be categorized into model-based algorithms and model-free algorithms which will be introduced in this section. In this section, we first briefly introduce DL in Subsection \ref{sec:dl}, then we detail DRL in Subsections \ref{sec:drl_model_free} and \ref{sec:drl_model_based} that correspond to model-free DRL algorithms and model-based DRL algorithms, respectively.

\subsection{Deep Learning: Review}
\label{sec:dl}
In this section, we review the most commonly used DL algorithms including autoencoders (AEs), deep belief networks (DBNs), convolutional neural networks (CNNs), recurrent neural networks (RNNs).

\subsubsection{Autoencoder}
Autoencoder is an unsupervised algorithm used for representation learning, such as feature selection or dimensionality reduction. An introduction to variational autoencoder (VAE) was given in \cite{doersch2016tutorial}. In general, VAE aims to learn a parametric latent variable model by maximizing the marginal log-likelihood of the training data.

\subsubsection{Deep belief network}
Deep belief networks and deep autoencoders are two commons unsupervised approaches that have been used to initialize the network instead of random initialization. While deep autoencoders are based on autoencoders which includes one visible inputs layer and one hidden layer, Deep Belief Networks is based on Restricted Boltzmann Machines which which contains a layer of input data and a layer of hidden units that learn to represent features that capture higher-order correlations in the data. 

\subsubsection{Multi-layer perceptron (MLP)}
Deep learning models, in simple words, are large and deep artificial neural networks. Let us consider the simplest possible neural network which is called ``\textbf{neuron}". A computational model of a single neuron is called a perceptron which consists of one or more inputs, a processor, and a single output.

Two main types of neural networks, \textit{i.e.}, convolutional neural networks and recurrent neural networks are introduced as follows.

\subsubsection{Convolutional neural network (CNN)}
Neural networks \cite{lecun1988theoretical} \cite{lecun1998efficient} are a special case of fully connected multi-layer perceptrons that implement weight sharing for processing data that has a known, grid-like topology (e.g. images). CNNs use the spatial correlation of the signal to constrain the architecture in a more sensible way. Their architecture, somewhat inspired by the biological visual system, possesses two key properties that make them extremely useful for image applications: spatially shared weights and spatial pooling. These kind of networks learn features that are shift-invariant, i.e., filters that are useful across the entire image (due to the fact that image statistics are stationary). The pooling layers are responsible for reducing the sensitivity of the output to slight input shift and distortions. Since 2012, one of the most notable results in Deep Learning is the use of convolutional neural networks to obtain a remarkable improvement in object recognition for ImageNet classification challenge \cite{deng2009imagenet} \cite{krizhevsky2012imagenet}. A typical convolutional network is composed of multiple stages. The output of each stage is made of a set of 2D arrays called feature maps. Each feature map is the outcome of one convolutional (and an optional pooling) filter applied over the full image.  

\subsubsection{Recurrent neural network (RNN)}
An RNN is an extremely powerful sequence model introduced in the early 1990s \cite{jordan1990long}. A typical RNN contains three parts, namely, sequential input data, hidden  state and sequential output data. RNNs make use of sequential information and perform the same task for every element of a sequence where the output is dependent on the previous computations. 

The difficulty of training an RNNs to capture long-term dependencies has been studied in \cite{Bengio1994}. To address the issue of learning long-term dependencies, \cite{Hochreiter_1997} proposed Long Short-Term Memory (LSTM), which is able to maintain a separate memory cell inside it that updates and exposes its content only when deemed necessary. Recently, a Gated Recurrent Unit (GRU) was proposed by \cite{Cho2014} to make each recurrent unit adaptively capture dependencies of different time scales. Like the LSTM unit, the GRU has gating units that modulate the flow of information inside the unit, but without having separate memory cells. 

\begin{figure*}[htbp]
	\centering \includegraphics[width=2\columnwidth]{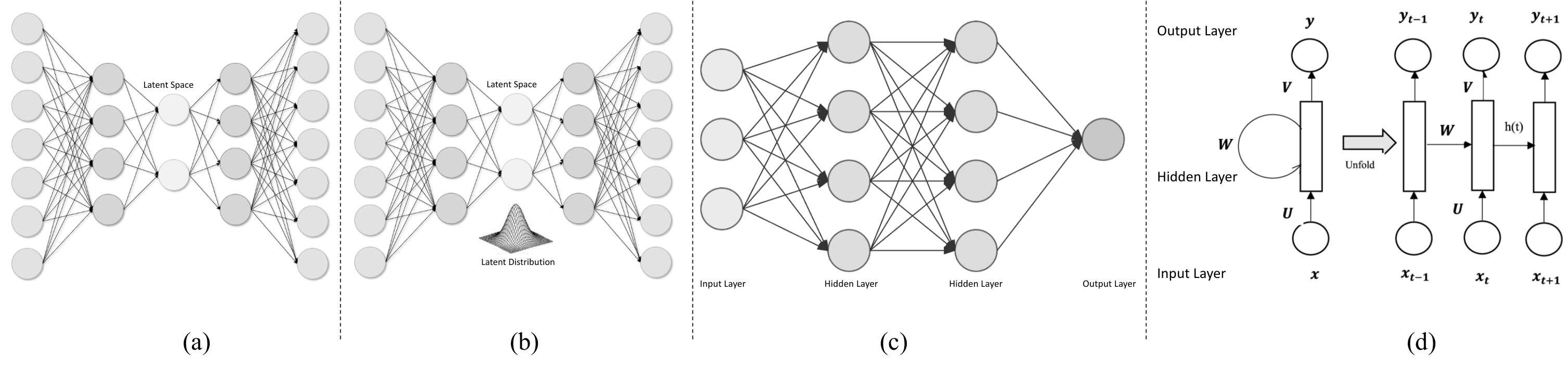}
		\caption[An illustration of various DL architectures]{An illustration of various DL architectures. (a): Autoencoder (AE); (b): Variational Autoencoder (VAE); (c): Convolutional Neural Network (CNN); (d): Recurrent Neural Network (RNN).}
	\label{fig:DL}
\end{figure*}

The visualization of various DL architecture networks is given in Fig. \ref{fig:DL}. 

\subsection{Model-free DRL algorithms}
\label{sec:drl_model_free}
There are three approaches, namely, value-based DRL methods,  policy gradient DRL methods and actor-critic DRL methods to implement model-free algorithms. The three approaches are detailed as follows.

\begin{figure}[htbp]
	\centering \includegraphics[width=0.9\columnwidth]{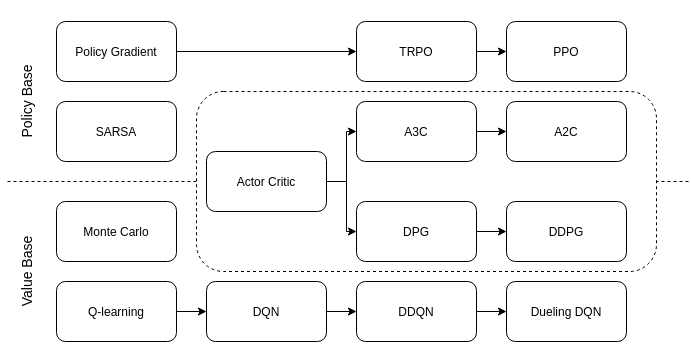}
	\caption{A roadmap of model-free reinforcement learning algorithms.}
	\label{fig:RL2}
\end{figure}

\subsubsection{Value-based DRL methods}
\begin{figure*}[htbp]
	\centering \includegraphics[width=2\columnwidth]{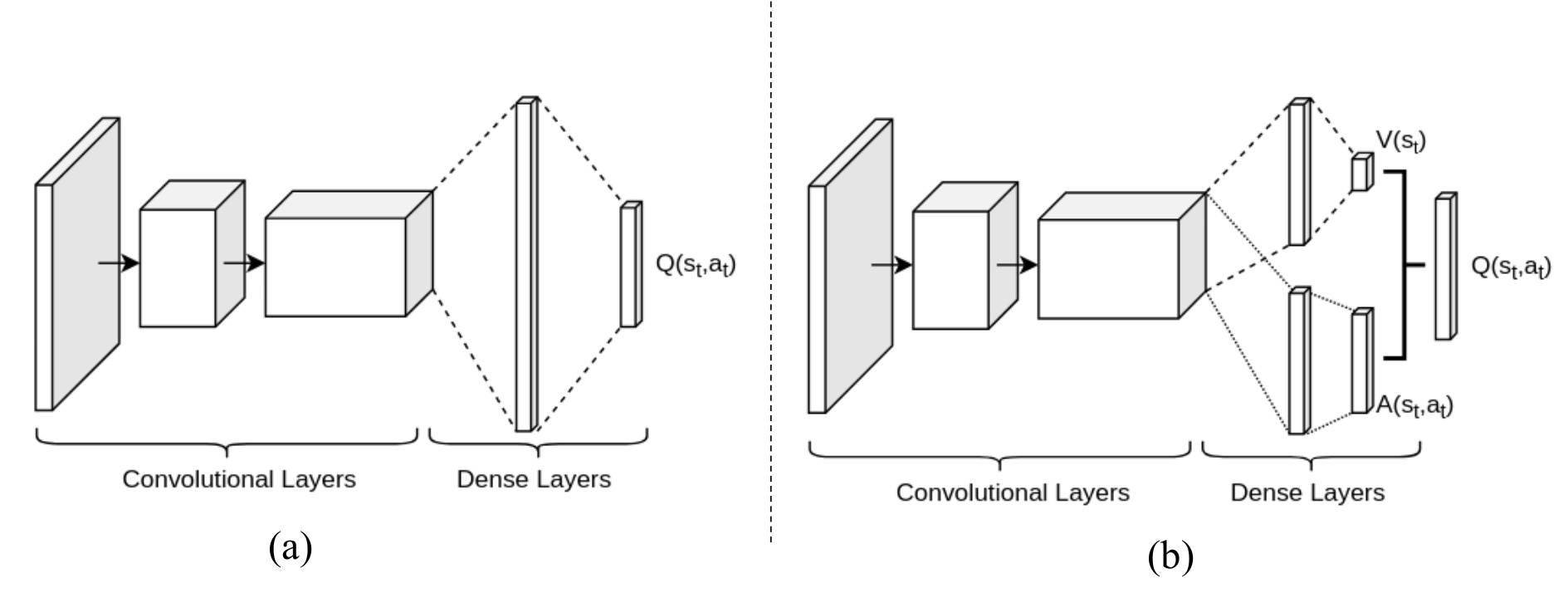}
		\caption{(a): Network structure of Deep Q-Network (DQN), where Q-values Q(s,a) are generated for all actions for a given state. (b): Network structure of Dueling DQN, where value function $V(s)$ and advantage function $A(s,a)$ are combined to predict Q-values $Q(s,a)$ for all actions for a given state. }
	\label{fig:dqn}
\end{figure*}

\underline{Deep Q-Learning Network (DQN).}
DQN \cite{mnih2015human} is the most famous DRL model which learns policies directly from high-dimensional inputs by a deep neural network as given in Fig.~\ref{fig:dqn}(a). Taking regression problem as an instance and letting $y$ denote the target of our regression task, the regression with input $(s, a)$, target $y(s, a)$ and the MSE loss function. The output $y$ and MSE loss are defined as in Eq.(\ref{eq:dqn}).
\begin{equation}
\begin{split}
    y(s_t,a_t) = R(s_t, s_{t+1}) + \gamma \max_{a_{t+1}}Q^*(s_{t_1}, a_{t+1}, \theta_t) \\
    \mathcal{L^{DQN}} = \mathcal{L}(y(s_t,a_t), Q^*(s_t, a_t, \theta_t)) \\ = ||y(s_t,a_t) - Q^*(s_t, a_t, \theta_t)||^2; \\
\end{split}
\label{eq:dqn}
\end{equation}
where $\theta$ is vector of parameters, $\theta \in \mathbb{R}^{|S||R|}$ and $s_{t+1}$ is a sample from $T(s_{t+1}| s_t, a_t)$ with input of $(s_t, a_t)$.

Minimizing the loss function yields a gradient descent step formula to update $\theta$ as follows:
\begin{equation}
    \theta_{t+1} = \theta_t -\alpha_t\frac{\mathcal{\partial L^{DQN}}}{\partial \theta}
\end{equation}

\underline{Double DQN.}
An improvement of DQN was introduced by Double DQN \cite{ddqn}.
One of the main limitation of DQN is that the values of $Q^*$ are tend to overestimation because of $max$ in Eq.~(\ref{eq:dqn}), $y(s,a) = R(s, s') + \gamma \max_{a'}Q^*(s', a', \theta)$ shifts Q-value estimation towards either to the actions with high reward or to the actions with overestimating approximation error. Double DQN is an improvement of DQN by combining double Q-learning \cite{hasselt2010double} with DQN to reduce observed overestimations with better performance.

The easiest but most expensive implementation of double DQN is to run two independent DQNs as follows:
\begin{equation}
\begin{split}
y_1 = R(s_t, s_{t+1}) + \gamma Q^*_1(s_{t+1}, \underset{a_{t+1}}{\arg\max}Q^*_2(s_{t+1},a_{t+1};\theta_2);\theta_1), \\
y_2 = R(s_t, s_{t+1}) + \gamma Q^*_2(s_{t+1}, \underset{a_{t+1}}{\arg\max}Q^*_1(s_{t+1},a_{t+1};\theta_1);\theta_2).
\end{split}
\end{equation}

\underline{Dueling DQN.}
In DQN, when the agent visits unfavourable state, instead of lowering its value $V^*$, it remembers only low pay-off by updating $Q^*$. In order to address this limitation, Dueling DQN \cite{wang2015dueling} incorporates approximation of $V^*$ explicitly in computational graph by introducing an advantage function as follows:
\begin{equation}
    A^{\pi}(s_t, a_t) = Q^{\pi}(s_t, a_t) - V^{\pi}(s_t).
\end{equation}
Therefore, Q-value is rewritten as
\begin{equation}
Q^{*}(s, a) = A^{*}(s, a) + V^{*}(s) \nonumber
\end{equation}
This implies that the feature map from DL is decomposed into with two parts corresponding to $V^*(v)$ and $A^*(s, a)$ as illustrated in Fig.\ref{fig:dqn}(b).
In practical implementation, Dueling DQN is formulated as follows:
\begin{equation}
\nonumber
    Q^{*}(s_t, a_t) =  V^{*}(s_t) + A^{*}(s_t, a_t) - mean_{a_{t+1}}A^{*}(s_t, a_{t+1}).
\end{equation}
Furthermore, to address the limitation of memory and imperfect information at each decision point, Deep Recurrent Q-Network (DRQN) \cite{Graves2013} employed RNN into DQN by replacing  the first fully-connected layer with a RNN. Multi-step DQN \cite{multi_step} is one of the most popular improvement of DQN by substituting one-step approximation with N-steps.

\noindent

\subsubsection{Policy gradient DRL methods}
\underline{Policy gradient theorem.}
Different from value-based DRL methods, policy gradient DRL optimizes the policy directly by optimizing the following objective function which is defined as a function of $\theta$:
\begin{equation}
    \mathcal{G}(\theta) = \mathbb{E}_{\mathcal{T}\sim\pi_{\theta}} \sum_{t=1}{\gamma^{t-1}R(s_{t-1}, s_t)} \rightarrow \max_{\theta}.
\label{eq:policy}
\end{equation}

For any MDP and differentiable policy $\pi_\theta$, the gradient of objective Eq.~(\ref{eq:policy}) is defined by policy gradient theorem \cite{gradient_policy} as follows:
\begin{equation}
    \bigtriangledown_\theta \mathcal{G}(\theta) = \mathbb{E}_{\mathcal{T}\sim\pi_{\theta}} \sum_{t=0}{\gamma^{t}Q^\pi(s_t,a_t)\bigtriangledown_\theta \text{log} \pi_{\theta}(a_t| s_t)}.
\label{eq:policy_gradient}
\end{equation}

\underline{REINFORCE.}
REINFORCE was introduced by \cite{williams1992simple} to approximately calculate the gradient in  Eq.~(\ref{eq:policy_gradient}) by using Monte-Carlo estimation. In REINFORCE approximate estimator, Eq.~(\ref{eq:policy_gradient}) is reformulated as:
\begin{equation}
    \bigtriangledown_\theta \mathcal{G}(\theta) \approx \sum_{\mathcal{T}}^N \sum_{t=0}{\gamma^{t}\bigtriangledown_\theta \text{log} \pi_{\theta}(a_t| s_t)(\sum_{t'=t}{\gamma^{t'-t}R(s_{t'}, s_{t'+1})})},
\label{reinforce}
\end{equation}
where $\mathcal{T}$ is trajectory distribution and defined in Eq.~(\ref{eq:T0}). Theoretically, REINFORCE can be straightforwardly applied into any parametric $\pi_{\theta}(a|s)$. However, it is impractical to use it because it is time consuming for convergence and there are local optima. Based on the observation that the convergence rate of stochastic gradient descent directly depends on the variance of gradient estimation, variance-reducing technique was proposed to address naive REINFORCE's limitations by adding a term that reduces the variance without affecting the expectation.

\subsubsection{Actor-critic DRL algorithm}
\begin{figure}[!h]
	\centering \includegraphics[width=0.9\columnwidth]{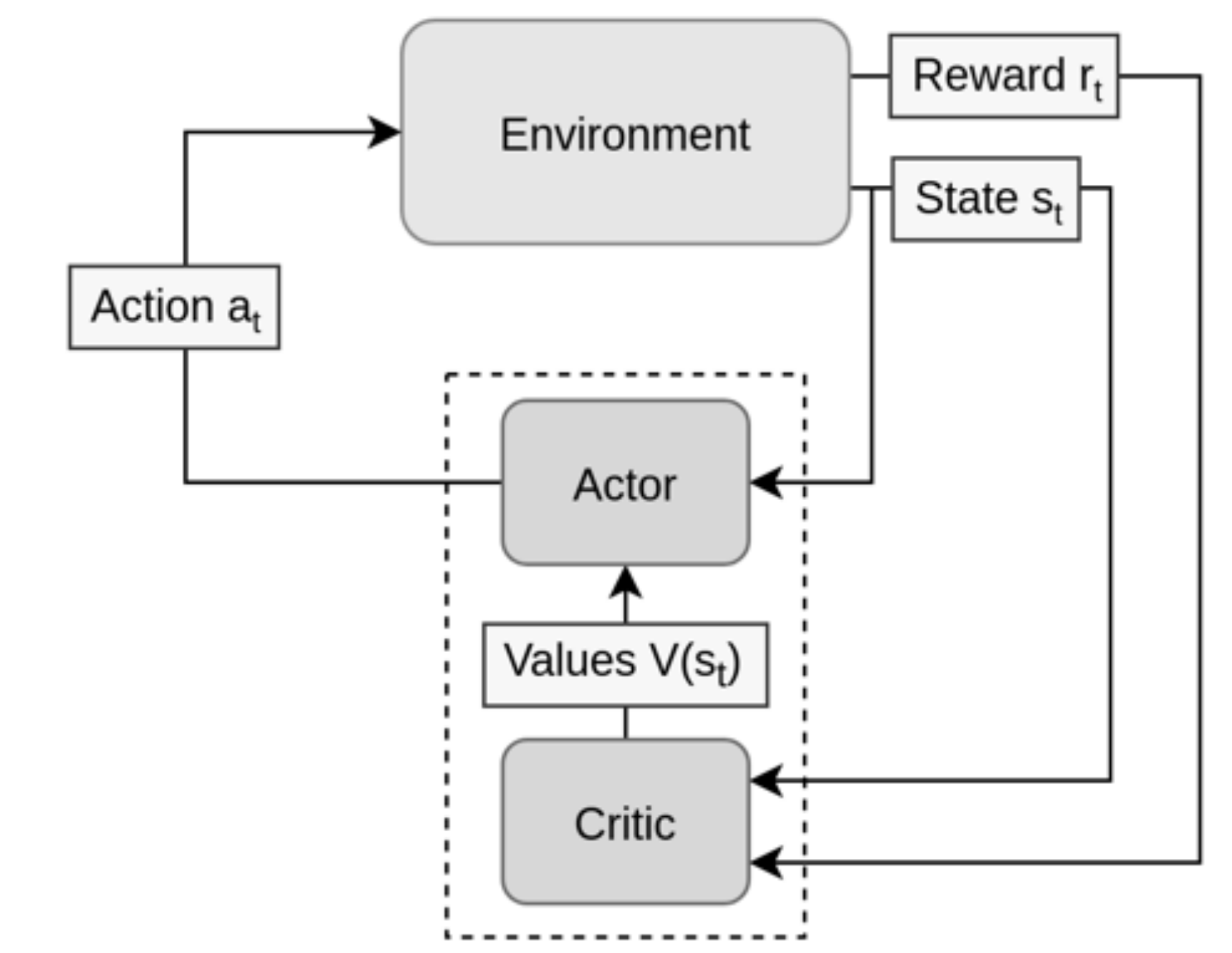}
		\caption{Flowchart showing the structure of actor critic algorithm. Action a, state s, reward r}
	\label{fig:a3c}
\end{figure}

Compared with value-based methods, policy gradient methods are better for continuous and stochastic environments and have a faster convergence. However, value-based methods are more sample efficient and steady. Lately, actor-critics \cite{konda2000actor} \cite{mnih2016asynchronous} was invented to take advantages from both value-based and policy gradient while limiting their drawbacks. Actor-critic architecture computes the policy gradient using a value-based critic function to estimate expected future reward. The principal idea of actor-critics is to divide the model in two parts: (i) computing an action based on a state and (ii) producing the $Q$ value of the action. As given in Fig.~\ref{fig:a3c}, the actor takes as input the state $s_t$ and outputs the best action $a_t$. It essentially controls how the agent behaves by learning the optimal policy (policy-based). The critic, on the other hand, evaluates the action by computing the value function (value based). The most basic actor-critic method (beyond the tabular case) is naive policy gradients (REINFORCE). The relationship between actor-critic is compared as a kid-mom relationship. The kid/actor explores the environment around with new actions while the mom/critic watches the kid and criticize/compliments. The kid then adjusts his behavior based on what his mom tells him. When the kid gets older, he is able to realize which action is bad/good.

\begin{figure}[!h]
	\centering \includegraphics[width=0.9\columnwidth]{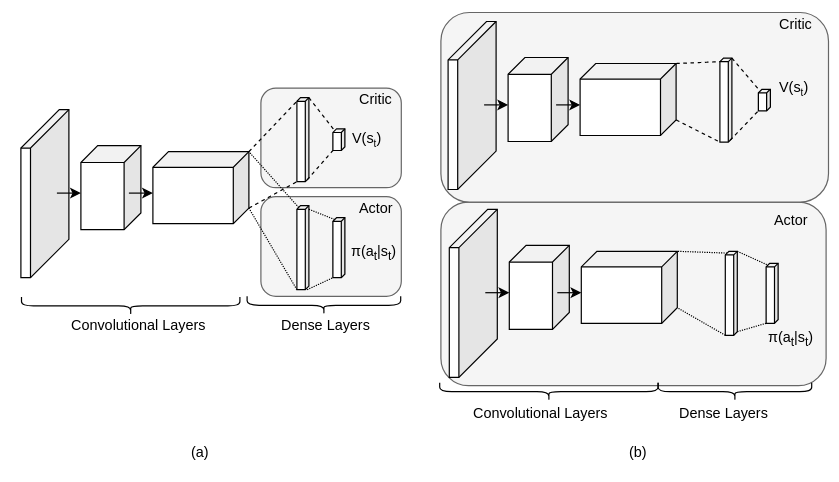}
		\caption{An illustration of Actor-Critic algorithm in two cases: sharing parameters (a) and not sharing parameters (b).}
	\label{fig:a3cn}
\end{figure}

\underline{Advantage actor-critic (A2C).}
Advantage actor-critic (A2C) \cite{a3c} consists of two neural networks, \textit{i.e.}, an actor network $\pi_\theta(a_t|s_t)$ representing for policy and a critic network $V^\pi_\omega$ with parameters $\omega$ approximately estimating actor’s performance.

At time step $t$, the A2C algorithm can be implemented as following steps:
 \begin{itemize}
     \item  Step 1: Compute advantage function:
            \begin{equation}
                A^\pi(s_t, a_t) = R(s_t, s_{t+1})+  \gamma V^\pi_\omega(s_{t+1}) - V^\pi_\omega(s_t)
            \label{eq:A2C_step1}
            \end{equation}
     \item Step 2: Compute target:
            \begin{equation}
                y = R(s_t, s_{t+1}) + \gamma V^\pi_\omega(s_{t+1})
            \label{eq:A2C_step2}
            \end{equation}
     \item Step 3: Compute critic loss with MSE loss:
     \begin{equation}
     \mathcal{L} = \frac{1}{B}\sum_T||y - V^\pi(s_t))||^2
     \end{equation}
     , where $B$ is batch size and $V^\pi(s_t)$ is defined by:
      \begin{equation}
      \begin{split}
        V^\pi(s_t) & = \mathbb{E}_{a_t\sim \pi( a_t| s_t)}\mathbb{E}_{s_{t+1}\sim T(s_{t+1}|a_t,s_t)}(R(s_t, s_{t+1}) \\ & + \gamma V^\pi(s_{t+1}))
    \end{split}
    \label{eq:A2C_V}
    \end{equation}
     \item Step 4: Compute critic gradient:
     \begin{equation}
     \bigtriangledown^{critic} = \frac{\partial \mathcal{L}}{\partial \omega}
     \end{equation}
     \item Step 5: Compute actor gradient:
     \begin{equation}
     \bigtriangledown^{actor} = \frac{1}{B}\sum_T{\bigtriangledown_\theta \text{log} \pi(a_t|s_t)A^\pi(s_t, a_t)}
     \end{equation}
 \end{itemize}

\underline{Asynchronous advantage actor critic (A3C).}
Beside A2C, asynchronous advantage actor critic (A3C) \cite{a3c} is an another strategy to implement an actor critic agent. To meet memory efficiency,  A3C asynchronously executes different agents in parallel on multiple instances of the environment instead of experience replay as in A2C.
Because of the asynchronous nature of A3C, some worker works with older values of the parameters and hence the aggregating update is not optimal. On the other hand, A2C synchronously updates the global network. A2C waits until all workers finished their training and calculated their gradients to average them, to update the global network.

In order to overcome the limitation of speed,  \cite{ga3c} proposed GA3C which achieves a significant speed up compared to the original CPU implementation. To more effectively train A3C, \cite{ftfe} proposed FFE which forces on random exploration at the right time during a training episode, which leads to improved training performance.

The structure of an actor-critic algorithm can be divided into two types, depending on whether or not parameter sharing is involved, as illustrated in Fig.\ref{fig:a3cn}.

\subsection{Model-based algorithms} \label{sec:drl_model_based}
We have discussed so far model-free methods including the value-based approach and policy gradient approach. In this section, we focus on the model-based approach, which deals with the dynamics of the environment by learning a transition model that allows for simulation of the environment without interacting with the environment directly. In contrast to model-free approaches, model-based approaches are learned from experience by a function approximation. Theoretically, no specific prior knowledge is required in model-based RL/DRL but incorporating prior knowledge can help faster convergence and better trained model, speed up training time as well as decreasing the required amount of training samples.  Also, it is difficult for model-based RL to directly use raw data with pixel as it is high dimensional. This is addressed in DRL by embedding the high-dimensional observations into a lower-dimensional space using autoencoders \cite{model_based_DRL}. Many DRL approaches have been based on scaling up prior work in RL to high-dimensional problems.  A good overview of model-based RL for high-dimensional problems can be found in \cite{plaat2020deep}, which partitions model-based DRL into three categories: explicit planning on given transitions, explicit planning on learned transitions, and end-to-end learning of both planning and transitions. In general, DRL targets at training DNNs to approximate the optimal policy $\pi^*$ together with optimal value functions $V^*$ and $Q^*$. In the following, we will cover the most common model-based DRL approaches including value function and policy search methods.

\subsubsection{Value function}
We start this category with DQN \cite{mnih2015human} which has been successfully applied to classic Atari and illustrated in Fig.\ref{fig:dqn}. DQN uses CNNs to deal with high dimensional state space to approximate the Q-value function.

\underline{Monte Carlo tree search (MCTS).} MCTS \cite{mtcs} is one of the most popular methods with look-ahead search and it is combined with DNN-based transition model to build a model-based DRL \cite{model-based-value_DRL0}. In this work, the learned transition model predicts the next frame and rewards one step ahead using the input of the last four frames of the agent’s first-person-view image and the current action. This model is then used by Monte Carlo tree search algorithm to plan the best sequence of actions for the agent to perform.

\underline{Value-targeted regression (VTR).} \cite{model-based-value_DRL} proposed model-based DRL for regret minimization. In their work, a set of models that are ‘consistent’ with the data collected is constructed at each episode. The consistency is defined as the total squared error, whereas the value function is determined by solving the optimistic planning problem with the constructed set of models.

\subsubsection{Policy search}
Policy search methods aim to directly find policies by means of gradient-free or gradient-based methods.

\underline{Model-ensemble trust-region policy optimization}\\
\underline{(ME-TRPO).} ME-TRPO \cite{model-based-value_DRL1} is mainly based on trust region policy optimization (TRPO) \cite{trpo} which imposes a trust region constraint on the policy to further stabilize learning.

\underline{Model-based meta policy optimization (MB-MPO).} MB-MPO \cite{model-based-policy_DRL} addresses the performance limitation of model-based DRL compared against model-free DRL when learning dynamics models. MB-MPO learns an ensemble of dynamics models and forms a policy that can quickly adapt to any model in the ensemble with one policy gradient step. As a result, the learned policy exhibits less model-bias without the need to behave conservatively.

A summary of both model-based and model-free DRL algorithms is given in Table \ref{tab:sum_drl}.

\begin{table*} \centering
\caption{Summary of model-based and model-free DRL algorithms consisting of value-based and policy gradient methods.}
\begin{tabular}{|l|l|l|}
\hline
    DRL Algorithms & Description & Category \\\hline
    DQN \cite{mnih2015human}& Deep Q Network & Value-based, Off-policy\\
    Double DQN \cite{ddqn} & Double Deep Q Network & Value-based, Off-policy \\
    Dueling DQN \cite{wang2015dueling} & Dueling Deep Q Network & Value-based, Off-policy \\
    MCTS \cite{model-based-value_DRL0} & Monte Carlo tree search & Valued-based, On-Policy \\
    UCRL-VTR\cite{model-based-value_DRL} & optimistic planning problem & Valued-based, On-Policy\\
    DDPG \cite{ddpg} & DQN with Deterministic Policy Gradient & Policy gradient, Off-policy \\
    TRPO \cite{trpo} & Trust Region Policy Optimization & Policy gradient, On-policy \\
    PPO \cite{ppo} & Proximal Policy Optimization & Policy gradient, On-policy \\
    ME-TRPO \cite{model-based-value_DRL1} & Model-Ensemble Trust-Region Policy Optimization & Policy gradient, On-policy \\
    MB-MPO \cite{model-based-policy_DRL} & Model-Based Meta- Policy-Optimization & Policy gradient, On-policy\\
    A3C \cite{a3c} & Asynchronous Advantage Actor Critic & Actor Critic, On-Policy \\
    A2C \cite {a3c} & Advantage Actor Critic & Actor Critic, On-Policy \\ \hline
\end{tabular}
\label{tab:sum_drl}
\end{table*}

\subsection{Useful techniques to train an agent}
In this section, we discuss some useful techniques that are used during training an agent.

\underline{Experience replay.}
Experience replay proposed by \cite{zha2019experience} is a useful part of off-policy learning. Experience replay is based on the fact that an agent can learn from some certain experiences (transitions , which may be rare but important) more than others (redundant transition or something already learned). By getting rid of as much information as possible from the past experiences, it removes the correlations in training data and reduces the oscillation of learning procedure.

\underline{Minibatch learning.}
Minibatch learning is a common technique that is used together with experience replay. Minibatch allows learning more than one training sample at each step, thus, it helps the learning process robust to outliers and noise.

\underline{Target Q-network freezing.}
As described in \cite{mnih2015human}, there are two networks in target Q-network freezing: one network interacts with the environment and another network plays a role of target network. The first network is used to generate target Q-values that are used to calculate losses. The weights of the second target network are fixed and slowly updated with the first network \cite{lillicrap2015continuous}.

\underline{Reward clipping.}
To keep the rewards in a reasonable scale and to ensure proper learning, they are clipped to a specific range (-1 ,1).

\section{DRL in Medical Imaging}
\label{sec:drl_mi}


We start with an exposition of the DRL formulation that is commonly used for parametric medical image analysis tasks such as landmark detection, image registration, and view plane localization. DRL also finds its use in other optimization tasks such as hyperparameter tuning, image augmentation selection, neural architecture search, etc., most of which share a common theme of non-differential optimization. Exhaustive grid search for these tasks is time-consuming and DRL is used to learn an efficient search policy. Finally, DRL is used in several miscellaneous topics.

\begin{figure}[htbp]
    \centering
    \includegraphics[width=0.9\columnwidth]{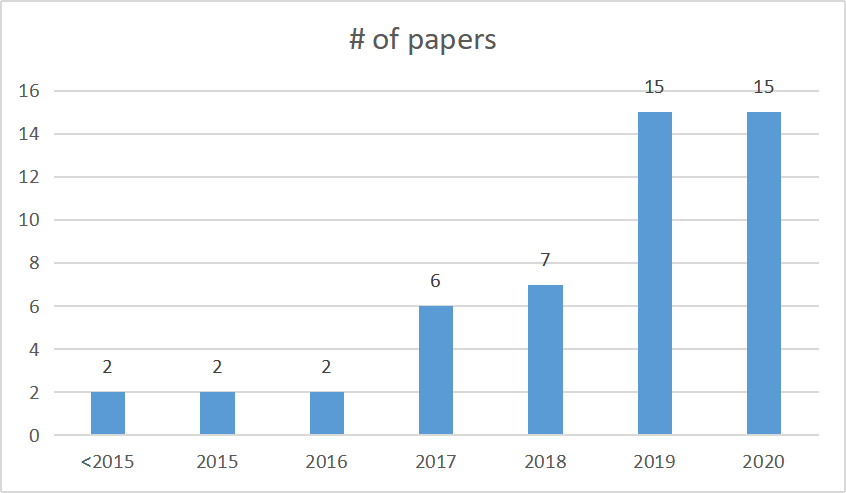}
    \caption{The number of DRL papers in medical imaging published every year.}
    \label{fig:papernum}
\end{figure}

Tables \ref{tab:ref}, \ref{tab:ref2}, and \ref{tab:ref3} contain a list of 49 references mostly published at top journals (such as IEEE Transactions Medical Imaging and Medical Image Analysis) and conferences (such as MICCAI). The list is by no means exhaustive. For each reference, we also provide the task with its concerned image modality and anatomy and offer some remarks when appropriate. Fig. \ref{fig:papernum} shows the number of DRL papers published every year, which clearly indicates a growing trend. In most of the listed papers, model-free learning algorithms are used.

\begin{sidewaystable*}[htbp]
\centering
\begin{tabular}{|l|c|l|l|}
\multicolumn{4}{c}{Category I: parameteric medical image analysis} \\
\hline
Task & Reference & Modality & Remarks\\
\hline
Landmark &\cite{ghesu2016artificial,ghesu2017robust,ghesu2018towards} & CT \& Ultrasound (US) & Multi-scale, using shape constraint\\
~~detection & \cite{alansary2019evaluating} & 3D US, fetal head & Multi-scale, multiple DRL strategies \\
& \cite{vlontzos2019multiple} & Brain MRI, fetal brain US & Multi-landmark det. via multi-agent RL \\
 & \cite{al2019partial} & CT \& MRI & Partial policy-based RL\\
 & \cite{zhang2020bottom} & 3D ultrasound & landmarking for mitral valve annulus modeling \\
  & \cite{Zhang_2020_localization} & MRI & Fetal pose detection \\
  & \cite{xu2017supervised} & 2D ultrasound & Supervised action classification\\
\hline
 Image & \cite{liao2017artificial} & CT and CBCT & Rigid registration \\
 ~~registration &  \cite{krebs2017robust} & MR prostate & Nonrigid registration\\
 & \cite{ma2017multimodal} & Depth to CT & 2 trans. \& 1 rotation parameters\\
 \hline
Object/lesion & \cite{maicas2017deep,maicas2019pre} & DCE-MRI & DRL for lesion bounding box detection \\
~~localization \&
& \cite{qaiser2019learning} & WSI microscopy & IHC scoring of HER2 \\
~~classification  & \cite{xu2019attention}& Breast histopathology & RL for selective attention\\
\hline
View plane & \cite{alansary2018automatic} & MR & Hierarchical action steps\\
~~localization & \cite{dou2019agent} & Fetal brain US & RL + warm start \& active termination \\
 & \cite{huang2020searching} & 3D ultrasound & Multi-plane localization \\
\hline
Plaque tracking & \cite{luo2019deep} & Intravascular OCT & 2 angles with 8 actions\\
\hline
Vessel extraction & \cite{zhang2018deep} & CT + MR & Tracing as sequential decision making\\
 & \cite{zhang2020branch} & Coronary CT angiography & Branch-aware Double DQN\\
\hline
\end{tabular}
\caption{A summary of references on parameteric medical image analysis using DRL.}
\label{tab:ref}
\end{sidewaystable*}

\begin{sidewaystable*}[htbp]
\centering
\begin{tabular}{|l|c|l|l|}
\multicolumn{4}{c}{Category II: Solving optimization using DRL} \\
\hline
Task & Reference & Modality & Remarks\\
\hline
Image and lesion & \cite{cheng2019adversarial} & Knee \& hip x-ray & Learning to mask an image as \\
~~classification &&&~~an attention map\\
&\cite{pesce2019learning} & Chest x-ray& Excluding unlabeled images w/o lesions \\
& \cite{akrout2019improving} & Visual skin image with Q's & RL agent to ask Q's \\
&& & ~~for improved performance\\ & \cite{ye2020synthetic} & Cervical \& lymph node & Synthetic sample selection via RL\\
& & ~~histopathology &\\
& \cite{wang2020auto} & Multimodal ultrasound & Auto-weighting for breast cancer classification \\
& \cite{ma2020attention} & MRI & Longitudinal Alzheimer’s disease analysis\\
\hline
Image & \cite{shokri2003using} & Various & traditional RL with \\
~~segmentation & \cite{sahba2006reinforcement} & & ~~limited modeling \\
& \cite{wang2013general} & & ~~such as thresholding\\
 & \cite{yang2019searching} & CT \& MRI, various &  Optimizing the DL training strategy\\
& \cite{bae2019resource} & Brain tumor, heart, prostate &  Neural architecture search \\
& \cite{yang2020deep} & 3D ultrasound & DQN-driven catheter segmentation\\
& \cite{qin2020automatic} & Kidney Tumor Segmentation & Automatic data augmentation via DRL\\
& \cite{liao2020iteratively} & Various & Interactive segmentation with multi-agent RL\\
 \hline
Image & \cite{zaech2019learning} & CBCT & Learning to avoid poor images\\
~acquisition& \cite{shen2018intelligent} & CT &  Tuning parameters for iterative recon.\\
& \cite{shen2020learning} & CT & Learning to scan\\
& \cite{pineda2020active} & MRI & Active k-space sampling\\
& \cite{li2020mri} & MRI & Pixel-wise operations using RL\\
\hline
Radiotherapy & \cite{shen2019intelligent} & & Tuning parameters for inverse \\
~~planning &&&~~treatment planning \\
\hline
\end{tabular}
\caption{A summary of references on solving optimization using DRL.}
\label{tab:ref2}
\end{sidewaystable*}

\begin{sidewaystable*}[htbp]
\centering
\begin{tabular}{|l|c|l|l|}
\multicolumn{4}{c}{Category III: Miscellaneous topics} \\
\hline
Task & Reference & Modality & Remarks\\
\hline
Video summarization & \cite{liu2020ultrasound} & Ultrasound video & Summarization using DRL\\
\hline
Surgical gesture & \cite{liu2018deep} & Surgical video & Small/large time step \\ \hline
Personalized & \cite{zhu2018group} & Smart device data & Group-driven RL\\
~~mHealth & & &\\ \hline
Model & \cite{neumann2015vito,neumann2016self} && Heart modeling \\
~~personalization & \cite{abdi2018muscle} &&Muscle  excitation estimation  \\
& \cite{joos2020reinforcement} && Musculoskeletal control  \\ \hline
\end{tabular}
\caption{A summary of references of miscellaneous topics that utilize DRL.}
\label{tab:ref3}
\end{sidewaystable*}

\subsection{DRL for parametric medical image analysis}

In many medical image analysis tasks, there are model parameters $\theta=[\theta_1, \theta_2, \ldots, \theta_n]$ to be estimated, given an image $I$. Table \ref{tab:task} exhibits a collection of common tasks and their associated model parameters. Currently, most model parameters are low dimensional.

\begin{table}[htbp]
    \centering
    \begin{tabular}{l|l}
         Task & Parameters \\ \hline
         2D landmark detection & $\theta=[x,y]$\\
         3D landmark detection & $\theta=[x,y,z]$\\
         Rigid 2D object detection & $\theta=[x,y,\alpha,s]$ \\
         Rigid 3D object detection & $\theta=[x,y,z,\alpha,\beta,\gamma, s]$ \\
         Rigid 2D/3D registration & $\theta=[x,y,z,\alpha,\beta,\gamma]$ \\
         View plane localization in 3D & $\theta=[a,b,c,d]$\\
         Others & $\theta$ depends on the task\\\hline
    \end{tabular}
    \caption{Common medical image analysis tasks and their associated model parameters. $x,y,z$ are for translation, $\alpha,\beta,\gamma$ for rotation, and $s$ for scaling.}
    \label{tab:task}
\end{table}

Below we first present a general DRL formulation for parametric medical image analysis and then proceed to cover each analysis task in separate subsection.

\subsubsection{Formulation}

To formulate a problem into the DRL framework, we have to define three key elements of DRL.

\underline{Action.} An action $a \in A$, where $A$ is the action space, is what the agent takes to interact with the environment, which is the image $I$.

One way of defining an action is to move each parameter, say the $i^{th}$ parameter, independently by $\pm\delta\theta_i$  while keeping the other parameters the same. The action space $A$ is given by:
\begin{equation}
    A = \{\pm\delta\theta_1, \pm\delta\theta_2, \ldots, \pm\delta\theta_n\}. \label{eq:actionspace}
\end{equation}
With this definition, the cardinality of the action space is $|A|=2n$.

The action space should be specified to guarantee the reachability, that is, starting an initial guess $\theta^0$, it is possible to reach an arbitrarily-valued parameter, say ${\hat \theta}=[{\hat \theta}_1, {\hat \theta}_2, \ldots, {\hat \theta}_n]$ . With the above definition, the reachability is trivially guaranteed, up to quantization error, by taking a series of actions: simply accumulating multiple steps of  $\pm\delta\theta_i$ to move the $i^{th}$ parameter by an amount of ${\hat \theta}_i-\theta^0_i$, and repeating this for each of the dimensions.

\underline{State.} The state is in regard to both the environment and the agent after all actions are taken so far.

Using the action space defined in (\ref{eq:actionspace}), the agent is at its state  $\theta_{t}$ after taking an action $a_t$:
\begin{equation}
    \theta_{t} = \theta_{t-1} + a_t = \theta_0 + \sum_{i=1}^t a_i.
\end{equation}
Note that the state of the environment is, an image (or image patch) ‘centered’ at $\theta_t$ denoted by $I[\theta_t]$.

\underline{Reward.} In general, the reward function should provide incentive signals when the target is hit or closer and penalize signals otherwise. Designing reward functions for reinforcement learning models is not easy. One design method is called inverse RL~\cite{inverseDRL} or ``apprenticeship learning", which learns a reward function that reproduces observed behaviors.

A commonly used reward function is given as below:
\begin{equation}
    R(s_t, s_{t-1}, a_t) = D(\theta_{t-1},{\hat \theta}) - D(\theta_{t},{\hat \theta}),\label{eq:reward}
\end{equation}
where $D(x,y)$ is a distance function that measures the difference between $x$ and $y$. If certain action reduces the difference, then a positive reward is obtained; otherwise, a negative reward is obtained.

To further intensify the effect of reward especially when the change in the difference is small, one can use
\begin{equation}
    R'(s_t, s_{t-1}, a_t) = sgn( R(s_t, s_{t-1}, a_t)),\label{eq:sgnreward}
\end{equation}
where $sgn(x)$ takes the sign of the value $x$. So, if certain action reduces the difference, then a positive reward $+1$ is obtained; otherwise, a negative reward $-1$ is obtained.

Once we have these three elements, we can invoke the DQL algorithm to trigger the learning process. Once the Q-function is learned, we can choose the action that maximizes the Q-function at each iteration.

It is clear that the search trajectory (or path) is implicitly related to the three elements. An alternative is to make the path explicit, that is, path supervision \cite{liao2017artificial,xu2017supervised}.
One path supervision approach is to guide the selection of the action that maximizes the reward in a greedy fashion for every iteration.
\begin{align}
    {\hat a}_t & = \arg \max_a R(s_t, s_{t-1}, a_t) \nonumber \\
    & = \arg \max_a D(\theta_{t-1},{\hat \theta}) - D(\theta_{t-1}+a,{\hat \theta}). \label{eq:pathsup}
\end{align}
This converts a reinforcement learning problem into supervised learning. With the pairs of  $(I[\theta_{t-1}],{\hat a}_t)$ forming training data, we can train supervised classification or regression functions.

\subsubsection{Landmark detection}

Medical landmarks are commonly used to represent distinct points in an image that likely coincide with anatomical structures. In clinical practices, landmarks play important roles in interpreting and navigating the image just like geographic landmarks that help travelers navigate the world. Also landmarks are used to derive measurements (e.g., width, length, size, etc.) of organs~\cite{xu2017supervised}, and to trigger subsequent, computationally intensive medical image analysis applications. In multi-modality image registration (such as PET-CT) or in registration
of follow-up scans, the fusion of multiple images can be initialized or guided by the positions of such
anatomical structures~\cite{johnson2002consistent,crum2004non}. In vessel centerline tracing, detected vessel bifurcations~\cite{liu2010search} can provide the start and end
points of certain vessels to enable fully-automated tracing~\cite{beck2010validation}. In organ segmentation, the center position of
an organ can provide the initial seed points to initiate segmentation algorithms~\cite{banik2009landmarking}. Landmark points situated on the organ surface, once detected, offer better initialization for segmentation~\cite{lay2013rapid}. In seminar reporting,
automatically found anatomical structures can be helpful in configuring the optimal intensity window for
display~\cite{pauly2011fast,lay2013rapid}, or offer the text tooltips for structures in the scan~\cite{seifert2010semantic}.

\underline{Artificial agent.}
In a series of papers, Ghesu et al. \cite{ghesu2016artificial,ghesu2017robust,ghesu2018towards} present a multi-scale approach for detecting anatomical landmarks in a 3D volume using an artificial agent. The landmark is represented as a 3D point and the actions include moving one-voxel step to the left, right, up, down, and forward and back. The reward function is given by (\ref{eq:reward}).

At each scale, a corresponding Q-function is learned to enable the agent to
effectively search for objects in the image, as opposed to scanning the volumetric space exhaustively. Per scale-space
theory, the system captures global context on coarse scale and local context on fine scale.
The search starts at the coarsest scale level, where the search model is trained for convergence
from any starting point in the image. On this scale level the field of view of the agent is very large with sufficient global context to ensure an effective navigation. Upon convergence,
the scale level is changed to the next level and the search continues. The process is repeated on the following scales until convergence
on the finest scale.

\begin{figure}
\centering
\includegraphics[width=\columnwidth]{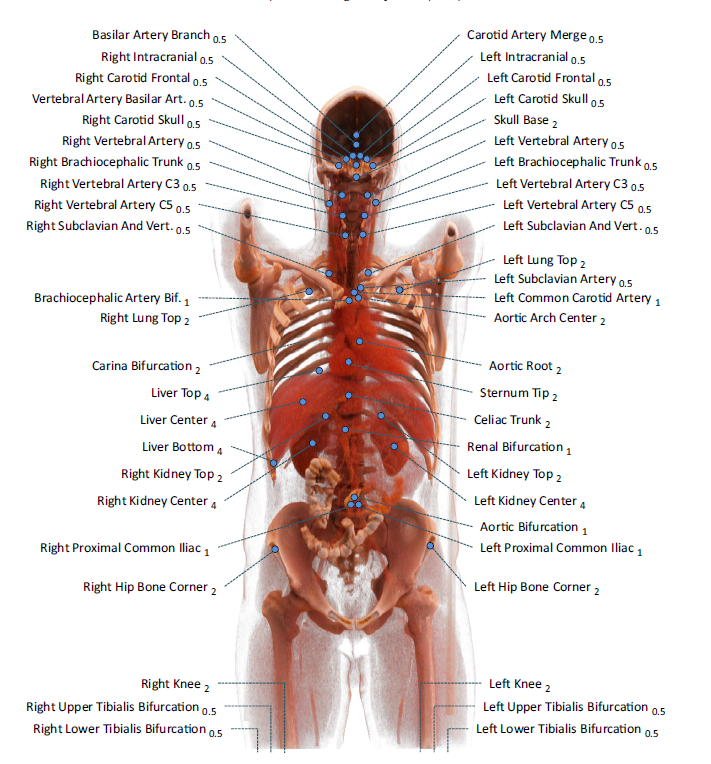}
\caption{The list of 49 anatomical landmarks. Courtesy of \cite{ghesu2018towards}.}
\label{F:lmk}
\end{figure}
The convergence criterion is met when trajectories converge on small, oscillatory-like cycles. Once such a cycle is identified at detection time, the search is stopped and the location is recorded as the detection result. An interesting finding is that, when searching for a landmark outside of the present scan, the search trajectory leaves the image space, signaling that the landmark is missing
from the field-of-view. To guarantee this consistent behavior, the system is trained by differently cropped images.

In addition, the constrained spatial distribution of anatomical landmarks using statistical shape modeling and robust estimation theory \cite{torr2000mlesac} is used to offer a probabilistic guarantee on the spatial coherence of the identified landmarks and to recognize if there are landmarks missing from the field-of-view. This shape fitting further makes the detection of landmarks more robust.

\begin{figure}
\centering
\includegraphics[width=\columnwidth]{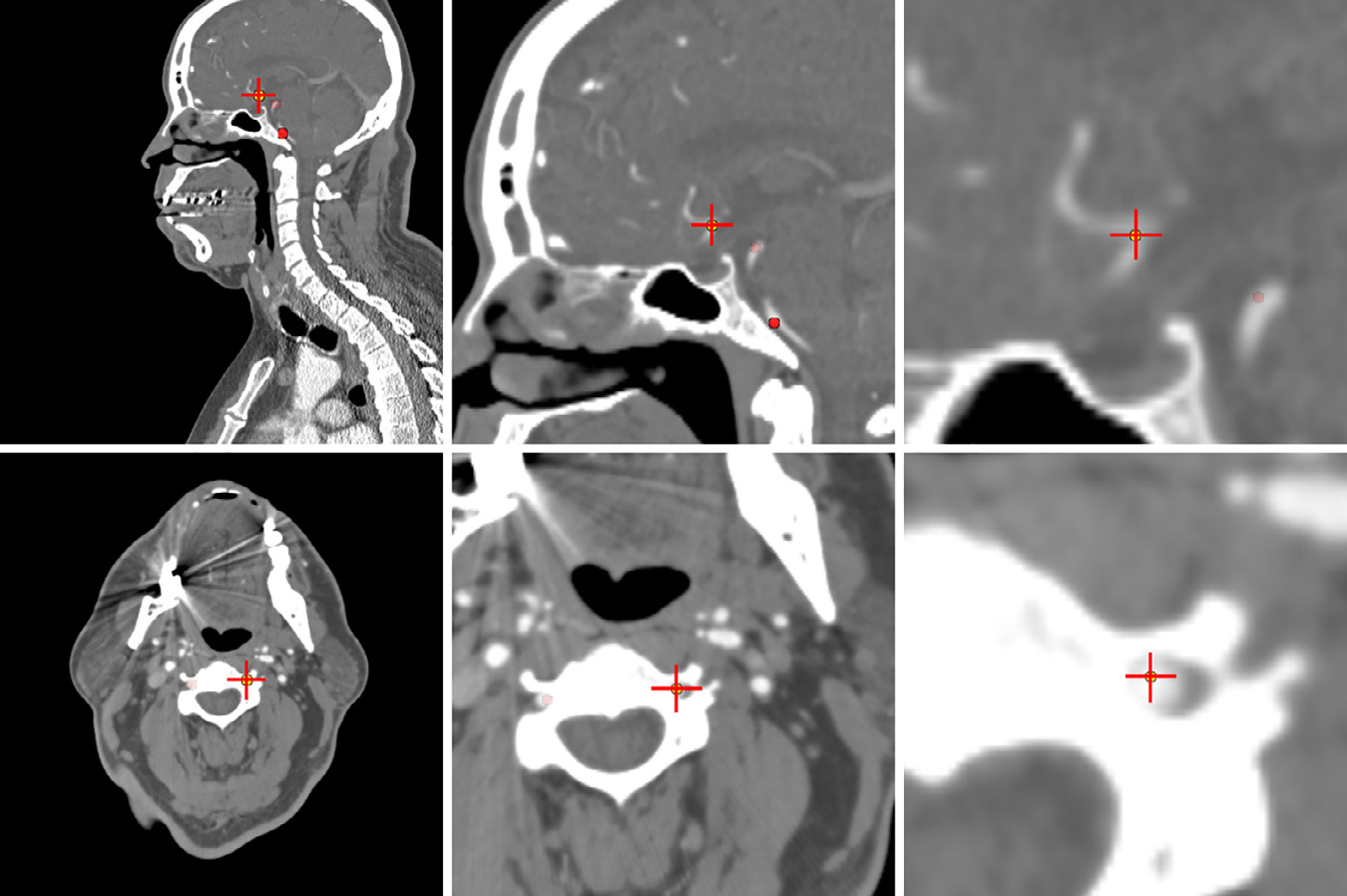}
\caption{Visualization of detection results of two vascular landmarks on 3 different levels from left to right. Courtesy of \cite{ghesu2018towards}.}
\label{F:vascularlmk}
\end{figure}

The proposed method is tested on detecting a cohort of 49 landmarks (see Figure \ref{F:lmk}) in a complete dataset of 5,043 3D-CT scans over 2,000 patients. When evaluating the detection performance, the landmarks 3cm within the image border are ignored. Perfect detection results with no false positive or negatives are reported. Figure \ref{F:vascularlmk} shows the detection results of two vascular landmarks on three different levels from left to right, which demonstrates the preciseness of the approach.

\cite{alansary2019evaluating} evaluate different reinforcement learning agents with different training strategies for detecting anatomical landmarks in 3D images. The specific training strategies include DQN, DDQN (Double DQN), Duel DQN, and Duel DDQN.  Also fixed- and multi-scale optimal path search strategies are compared. The finding is that the optimal DQN architecture for achieving the best performance depends on the environment.

\cite{vlontzos2019multiple} consider the interdependence between multiple landmarks as they are associated with the human anatomy. It is likely that localizing one landmark helps detect the other landmarks. They propose to train a set of multiple collaborative agents using reinforcement learning in order to detect multiple landmarks, instead of a naive approach that learns many separate agents, one agent for each landmark.
It is shown that the multi-agent RL achieves significantly better accuracy by reducing the detection error by 50\% on detecting 7-14 landmarks for three tasks, consumes fewer computational resources, and reduces the training time, when compared with the naive approach.

In \cite{al2019partial}, an RL agent is learned for landmark localization in 3D medical images, following the formulation in~\cite{ghesu2017robust}. However, an actor-critic approach is utilized to directly approximate the policy function
In addition, in order to speed up the learning and reach a more robust localization, multiple partial policies on different sub-action spaces are learned instead
of a single complex policy on the original action space. For a 3D landmark $(x,y,z)$, the action space is $A = \{\pm\delta x, \pm\delta y, \pm\delta z\}$; so it is natural to define three sub-action
spaces $A_k=\{\pm\delta k\}$ with $k \in \{x,y,z\}$
by projecting the actual action
space onto different Cartesian axes. Experiments on three datasets, namely 71 contrast-enhanced coronary CT angiography volumes with 8 landmarks, 150 cardiac CT volumes with a landmark of left atrial appendage (LAA) seed-point, and 18 MR spine images with 5 lumbar vertebra landmarks,  demonstrate that the proposed actor-critic approach with partial policies achieve robust and improved performances, compared to the conventional actor-critic and widely used deep Q-learning approach.

\cite{zhang2020bottom} propose a bottom-up approach for automatically building a mitral valve annulus model from 3D echocardiographic images in real time, in which the very first step is to automatically detect a few key landmarks associated with the above annulus model using the artificial agent~\cite{ghesu2017robust}.

\cite{Zhang_2020_localization} incorporate priors on physical structure of the fetal body to optimize multi-agent for detection of fetal landmarks. In this work, they use graph communication layers to improve the communication among agents based on a graph where each node represents a fetal body landmark. The proposed network architecture contains two parts corresponding to shared CNNs for feature extraction and graph communication networks to merge the information of correlated landmarks. Furthermore, the distance between agents and physical structures such as the fetal limbs is used as a reward. The evaluation is conduction on 19,816 3D BOLD MRI volumes acquired on a 3T Skyra scanner. The proposed method achieves an average detection accuracy of 87.3\% under a 10-mm threshold and 6.9mm as the mean error.

\underline{Supervised action classification.}
\cite{xu2017supervised} propose to approach landmark detection as image partitioning. This nontrivial approach is derived from path supervision.

\begin{figure}
\centering
\includegraphics[width=\columnwidth]{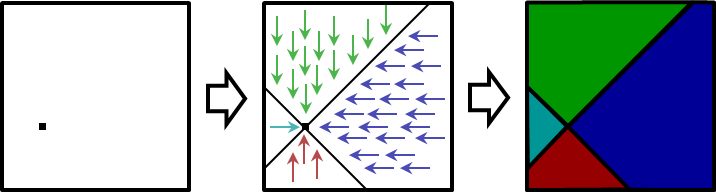}
\caption{The discrete action map representation.}
\label{F:map}
\end{figure}

Consider an agent that seeks an optimal action path from any location at $(x,y)$ towards a landmark $l=(\hat{x},\hat{y})$, which is composed of optimal action steps at pixels along the path on an image grid $\Omega$. In other words, at each pixel the agent is allowed to take an action $a$ with a unit movement $d_x^{(a)} \in \{-1,0,1\}$ and $d_y^{(a)} \in\{-1,0,1\}$. With the constraint of
$\|d_x^{(a)}\|^2+\|d_y^{(a)}\|^2=1$,
we basically allow four possible action types $a\in \{0,1,2,3\}$:
\begin{eqnarray*}
UP: & (d_x^{(0)}=~0,~~d_y^{(0)}=-1),\\
RIGHT: & (d_x^{(1)}=~1,~~d_y^{(1)}=~0),\\
DOWN: & (d_x^{(2)}=~0,~~d_y^{(2)}=~1),\\
LEFT: & (d_x^{(3)}=-1,~~d_y^{(3)}=~0).
\end{eqnarray*}
The optimal action step $\hat{a}$ is selected as the one with minimal Euclidean distance to the landmark $l$ after its associated movement,
\begin{equation}
\hat{a}=\arg\min_{a} \sqrt{(x-\hat{x}+d_x^{(a)})^2+(y-\hat{y}+d_y^{(a)})^2}. \label{eq:a}
\end{equation}
Simple derivations show that the selection of $\hat{a}$ falls into four regions (one for each action type), where the regions are partitioned by two lines with slopes of $\pm 1$ crossing the landmark (Figure~\ref{F:map}):
\begin{eqnarray*}
y=x+(\hat{y}-\hat{x} ),~~
y=-x+(\hat{x}+\hat{y}).
\end{eqnarray*}
This generates a discrete action map $a(x,y)$ that represents the pixel-wise optimal action step moving toward the target landmark location.

During training to estimate the action map for a given image, a fully convolutional neural network, called a deep image-to-image network (DI2IN), can be employed given its efficient sampling scheme and large receptive field for comprehensive feature learning. During testing, the landmark location needs to be derived from the estimated action map.
To this, an aggregate approach is proposed. 
With the output action map $A(x,y)$ from DI2IN, the estimated landmark location coordinates $(x',y')$ are determined by maximizing an objective function $C(\cdot)$ summed up with that of each action type $C_a (\cdot)$.
\begin{equation}
(x',y')= \arg\max_{(x,y)} C(x,y) = \arg \max_{(x,y)} \sum_{a} C_{a} (x,y),
\end{equation}
where the action-wise objective function at pixel $(x,y)$ is aggregated by the pixels with that specific action on the same row or column, specifically
\begin{equation}
C_a (x,y)=
\begin{cases}
   d_x^{(a)}\{
   \sum_i (2~\pi[i\ge x]-1)\pi[A(i,y)==a]
  \} \\
   ~~~~~~~~~~~~~~~~~~~~
   \text{if } \|d_x^{(a)}\|=1,\\
   d_y^{(a)}\{
   \sum_j (2~\pi[j\ge y]-1) \pi[A(x,j)==a]\} \\
   ~~~~~~~~~~~~~~~~~~~~
   \text{if }  \|d_y^{(a)}\|=1,
  \end{cases}
\end{equation}
where $\pi[.]$ is an indicator function.
Such aggregation enables robust location coordinate derivation even with suboptimal action map from the DI2IN output.

In experiments on detecting landmarks from a cardiac or obstetric ultrasound image in two datasets with 1,353 and 1,642 patients, respectively, it is demonstrated that the proposed approach achieves the best results when compared with state-of-the-art approaches that include the artificial agent.

\subsubsection{Image registration}

Robust image registration in medical imaging is essential for comparison or fusion of images, acquired from various perspectives, in different modalities or at different times. In terms of modeling the registration, there are two ways: rigid and non-rigid.

\underline{Rigid registration.}
Rigid registration is fully specified by a few number of transformation parameters. For example, a 3D rigid registration typically has 6 parameters to optimize. Traditionally, image registration is solved by optimizing an image matching metric such as normalized correlation coefficient or mutual information as a cost function, which is difficult due to the non-convex nature of the matching problem.

\cite{liao2017artificial} propose an artificial agent to perform image registration. It casts the image registration problem
as a process of finding the best sequence of motion actions (\textit{e.g.,} up, down, left, right, etc.) that
yields the desired image registration parameter.
The input to the agent is the 3D raw image data and the current estimate of image registration parameter, and the output of the agent, which is modeled using a deep convolutional neural network, is the next optimal action.
Further, it utilizes the path supervision approach to supervise the end-to-end training. Since the agent is learnt, it avoids the issue of current approaches that are often customized to a specific problem and sensitive to image quality and artifacts.

\begin{figure}
\centering
\includegraphics[width=\columnwidth]{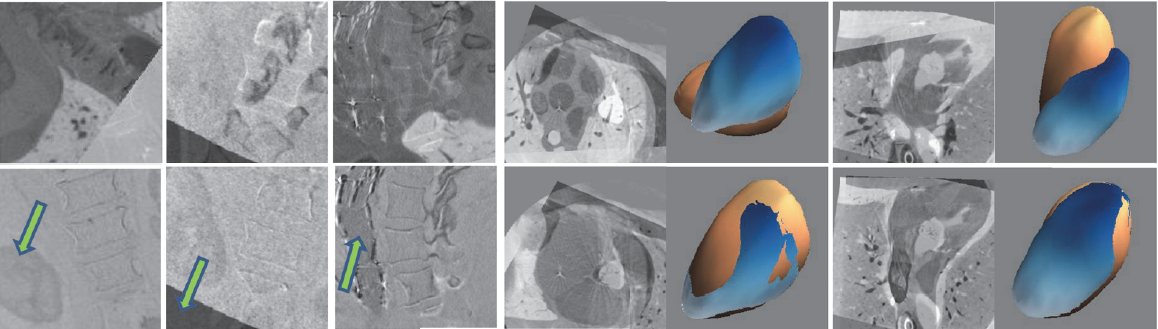}
\caption{Registration examples shown as the difference between the reference and floating images, before (upper row) and after (lower row) registration. The mesh overlay before and after registration is shown for cardiac use case for improved visualization. Picture courtesy of \cite{liao2017artificial}.}
\label{F:regis}
\end{figure}

In experiments, the proposed approach is evaluated on two datasets: spine (87 pairs of images) and heart (97 pairs of images). In the first dataset of aligning abdominal spine CT and CBCT, the main challenging lies in that CT has a much larger FOV than CBCT, leading to many local optima in the registration space due to the repetitive nature of the spine. In the second dataset of registering cardiac CT and CBCT (as in Figure~\ref{F:regis}), the main challenge lies in the poor
quality of CBCT with severe streaking artifacts and weak
soft tissue contrast at the boundary of the epicardium.
On both datasets, the artificial agent outperforms several state-of-art registration methods by a large margin in terms of both accuracy and robustness.

Similarly, \cite{ma2017multimodal} use the artificial agent to register a 2.5D depth image and a 3D CT. Different from \cite{liao2017artificial}, it uses dueling DQN to learn the Q function instead of path supervision. Further, although it involves a six degree-of-freedom transformation, the search space is simplified into two translations and one rotation as the rest of the transformation can be determined/inferred through the sensor calibration process together with the
depth sensor readings. It also invokes orthographic projection to generate 2D images that are fed into the Q function. Quantitative evaluations are conducted on 1788 pairs of CT and depth images from real clinical setting, with 800 as training. The proposed method achieves the state-of-the-art performance, when compared with several approaches including \cite{ghesu2016artificial}.

\underline{Non-rigid registration.}
When rigid transformation is insufficient to describe the transformation between two images, a non-rigid registration comes into play, which has more than 6 parameters in 3D to optimize, depending on the class of non-rigid registration.

\cite{krebs2017robust} extend the artificial agent approach to handle non-rigid registration. In particular, the parametric space of a statistical deformation model for an organ-centered registration of MR prostate images is explored. There are $m=15$ PCA modes in 2-D and $m=25$ modes in 3-D kept to model the prostate deformation, with $2 \times m$ actions are defined.

To tackle the difficulty of obtaining trustworthy ground-truth deformation fields, \cite{krebs2017robust} proceed with a large number of synthetically deformed image pairs derived from only a small number of inter-subject pairs. Note that the extracted ground truth reaches a median DICE coefficients of 0.96 in 2-D and 0.88 in 3-D. The Q function is then learned.

The algorithm is tested on inter-subject registration of prostate MR data (41 3D volume in total with 8 for testing, resulting in 56 inter-subject pairs). For the 2D experiment, the middle slice of each volume is utilized. Before the non-rigid registration, the initial translation registration is performed using the elastix approach \cite{klein2010elastix} by registering each of the test images to an arbitrarily chosen template from the training database. The final registration result reaches a median DICE score of 0.88 in 2-D and 0.76 in 3-D, both better than competing state-of-the-art registration algorithms.

\subsubsection{Object/lesion localization and detection}

DRL is also leveraged to detection objects \cite{jie2016tree}. \cite{maicas2017deep} present such an approach for detecting breast lesions from dynamic contrast-enhanced magnetic resonance imaging (DCE-MRI).

The bounding box for a 3D lesion is defined as ${\bf b}=[b_x,b_y,b_z,b_w,b_h,b_d]$ and the actions are defined as $\{l_x^+, l_x^-,l_y^+, l_y^-,l_z^+, l_z^-,s^+,s^-,w\}$, where $l,s,w$ represent translation, scale and trigger actions, with the subscripts $x,y,z$ denoting the horizontal, vertical or depth translation, and superscripts $+,-$ meaning positive
or negative translation and up or down scaling. The signed reward function is used. DQN is learned based on the ResNet architecture.

\begin{figure}
\centering
\includegraphics[width=\columnwidth]{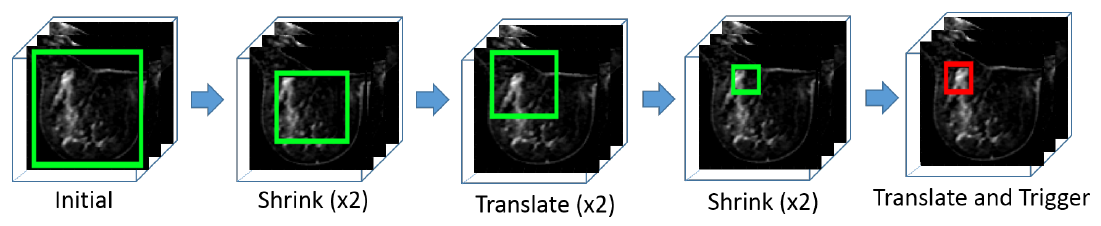}
\caption{The illustration of the detection process, with the learnt DRL agent outputting a series of allowable actions to realize final detection of a 3D lesion. Picture courtesy of \cite{maicas2017deep}.}
\label{fig:lesion}
\end{figure}

Experiments are conducted on DCE-MRI volumes from 117 patients. The training set contains 58 patients annotated with 72 lesions, and the testing set has 59 patients and 69 lesions. Results show a similar accuracy to state of the art approaches, but with significantly reduced detection time.

\cite{pesce2019learning} study how to localize pulmonary lesions in a chest radiograph. In one of the proposed methods,
a recurrent attention model with annotation feedback (RAMAF) is learned using RL to
observe a short sequence of image patches. The classification score is used as a reward signal, which penalises the exploration of areas that are unlikely to contain nodules and encourages the learning of a policy that maximises the conditional probability of the true label given a series of image patches within the radiographs.

\begin{figure}
\centering
\includegraphics[width=\columnwidth]{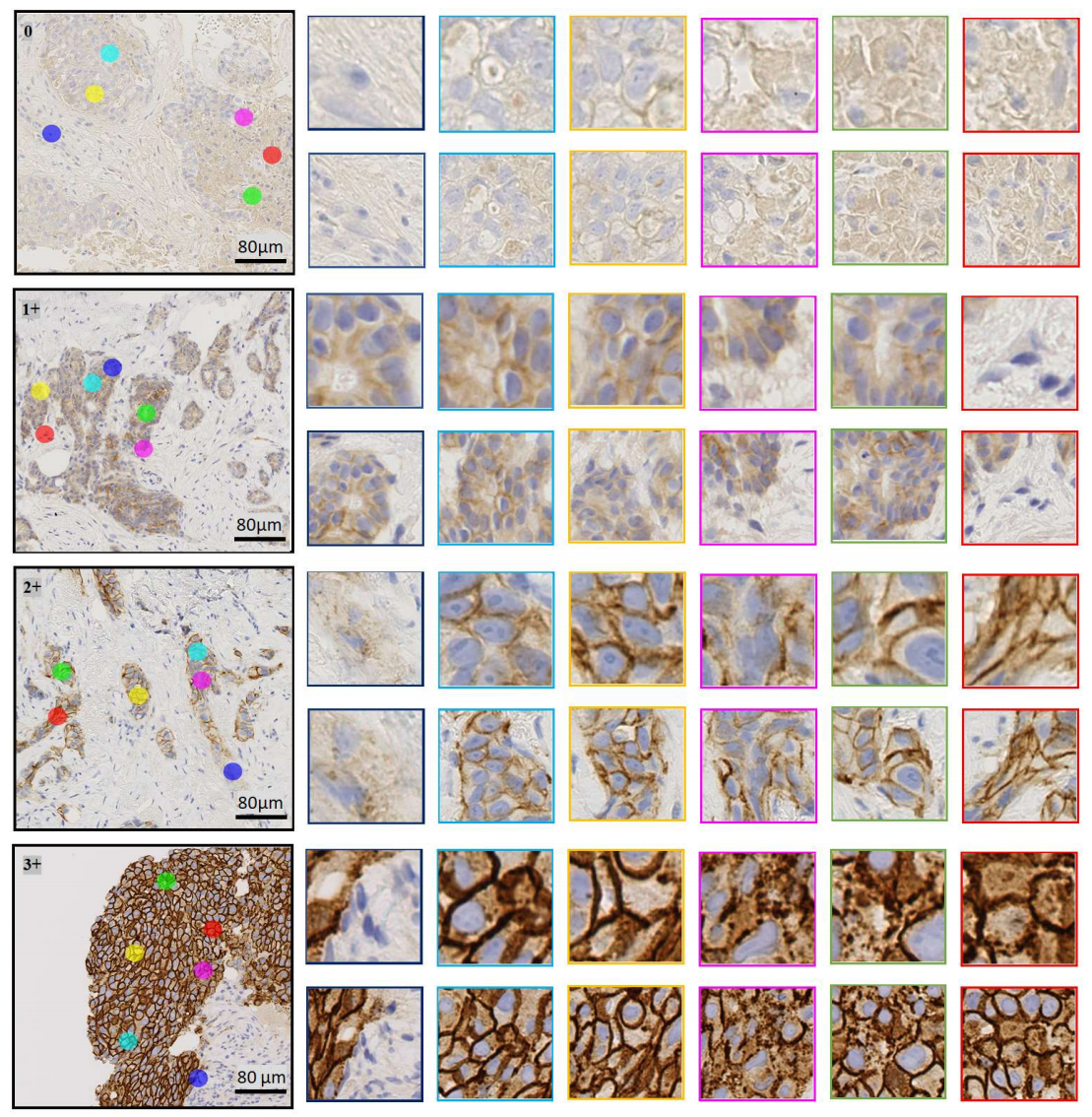}
\caption{Example of four image tiles with selected regions-of-interest (ROIs) predicted by~\cite{bae2019resource}, for each HER2 score (0-3+), respectively. The first column shows the input images and colored disks shows the predicted locations. The remaining columns show the selected regions at 40$\times$ and 20$\times$ around the selected locations. The first selected region is shown with blue bounding boxes and the last selected region is shown with red bounding boxes. Picture courtesy of~\cite{bae2019resource}.}
\label{F:wsi}
\end{figure}

In \cite{qaiser2019learning}, a sequential learning task is formulated to
estimate from a giga-pixel whole slide image (WSI) the immunohistochemical (IHC) scoring of human epidermal growth factor receptor 2 (HER2) on invasive breast cancer (BC), which is a significant predictive and prognostic marker. To solve this task, DRL is employed to learn a
parameterized policy to identify diagnostically relevant regions of interest (ROIs) based on current inputs, which are comprised of two image patches cropped at 40$\times$ and 20$\times$ magnification levels. The selected ROIs are processed by a CNN for HER2 scores. This avoids the need to process all the sub-image patches of a given tile and saves a large of amount of computations. Refer to Figure~\ref{F:wsi}
for some illustrative results of HER2 scoring.

\cite{xu2019attention} take the computational challenge of breast cancer classification from a histopathological image. Due to the large size of a histopathological image, pathologists in clinical diagnosis first find an abnormal region and then investigate the detail within the region. Such a human attention mechanism inspires an attention-based deep learning approach. It consists of two networks for selection and classification tasks separately. The selection network is trained using DRL, which outputs a soft decision about whether the cropped patch is necessary for classification. These selected patches are used to train the classification network, which in turn provides feedback
to the selection network to update its selection policy. Such a co-evolution training strategy enables fast convergence and high classification accuracy. Evaluation based on a public breast cancer histopathological image database of 7,909 images and eight subclasses of breast cancers from 82 patients (58 malignant and 24 benign) demonstrates about 98\% classification accuracy while only taking 50\% of the training time of the previous hard-attention approach.


\subsubsection{View plane localization}

\cite{alansary2018automatic} propose to use DRL to detect canonical view planes in MR brain and cardiac volumes. A plane in 3D $ax+by+cz+d=0$ is parameterized by a 4D vector $[a,b,c,d]$. The eight actions are defined as $\{\pm \delta_{\theta_x},\pm \delta_{\theta_y}, \pm \delta_{\theta_z},\delta_{d},\}$, which update the plane parameters as $a=cos(\theta_x+\delta_{\theta_x})$, $b=cos(\theta_y+\delta_{\theta_y})$,  $c=cos(\theta_z+\delta_{\theta_z})$, and $d=d+\delta_d$. The signed reward function is used. Further a multi-scale strategy is utilized, with the action steps are refined a coarse-to-fine fashion.

\begin{figure}
\centering
\includegraphics[width=0.2\columnwidth]{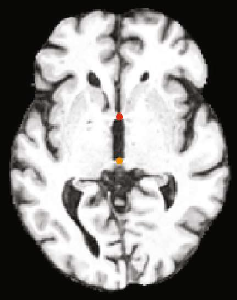}(a)
\includegraphics[width=0.3\columnwidth]{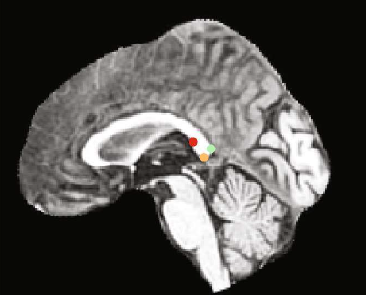}(b)
\includegraphics[width=0.25\columnwidth]{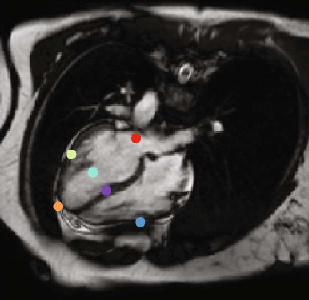}(c)
\caption{The viewing planes for detection: (a) Brain axial ACPC plane,  (b) Brain mid saggital plane (c) Cardiac apical four chamber plane. The landmarks are visualized for better definition of the plane and used for error calculation. Picture courtesy of \cite{alansary2018automatic}.}
\label{fig:planes}
\end{figure}

The experiments are based on 382 brain MR volumes (isotropic 1mm) and 455 short-axis cardiac MR volumes (1.25$\times$1.25$\times$2mm$^3$). Figure \ref{fig:planes} visualize the viewing planes to be detected. The specific Q-learning strategies include DQN, DDQN (Double DQN), Duel DQN, and Duel DDQN. The detection of the ACPC and mid-saggittal planes reaches an error less than 2mm and the detection of the apical four chamber plane reaches an error around 5mm.

\cite{dou2019agent} study how to use a DRL agent to localize two standard planes of transthalamic (TT) and transcerebellar (TC) positions in a 3D ultrasound volume of fetal head. The plane parameterization, action space, and reward function are defined in a similar manner to \cite{alansary2018automatic}. To ease the localization, they propose to augment the agent with a warm start module for better initialization and an active termination module for drift prevention. Based on their extensive validation on in-house datasets of 430 prenatal US volumes, the proposed approach improves both the accuracy and efficiency of the localization system.

\cite{Huang_2020_USplane} localize multiple uterine standard planes in 3D ultrasound simultaneously by a multi-agent DRL, which is equipped by one-shot neural architecture search (NAS) module. In this work, gradient-based search using a differentiable architecture sampler (GDAS) is employed to accelerate and stabilize the training process. Furthermore, to improve the system robustness against the noisy environment, a landmark-aware alignment model is utilized. The spatial relationship among standard planes is learnt by a recurrent neural network (RNN). They conduct the experiment on in-house dataset of 683 volumes which show that multiple agents with recurrent network obtain the best performance.


\subsubsection{Plaque tracking}

\begin{figure}[htbp]
\centering
\includegraphics[width=\columnwidth]{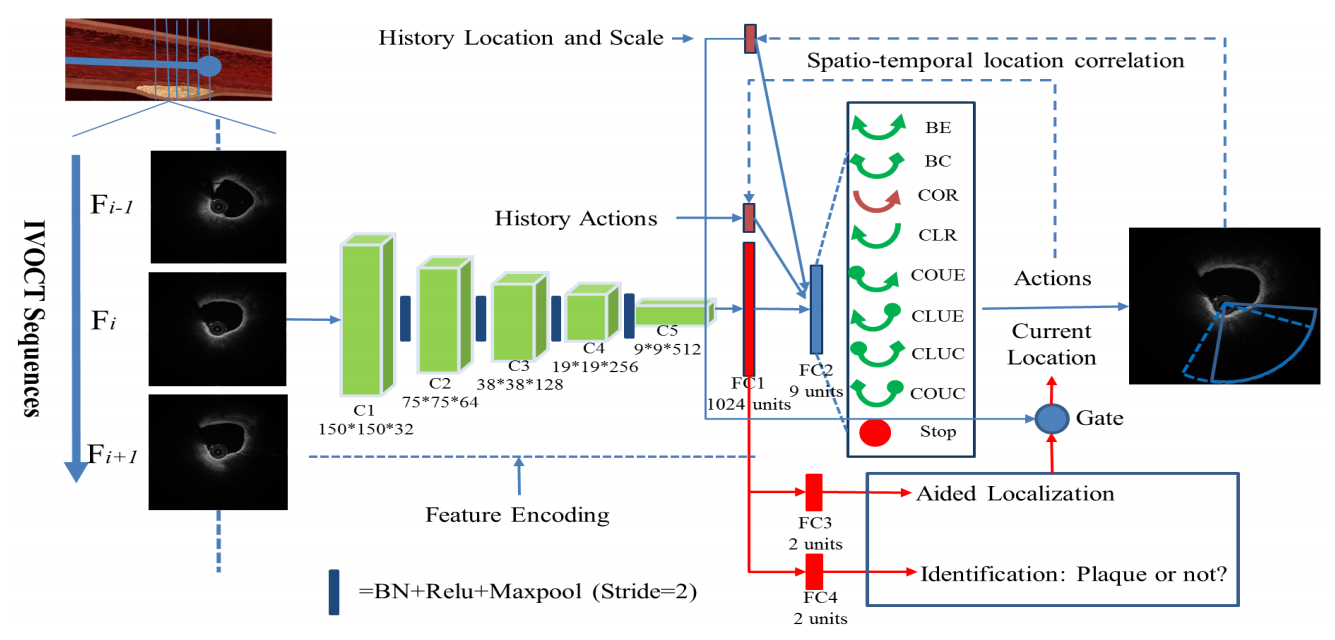}
\caption{The DRL framework is proposed to leverage the spatiotemporal information to achieve continuous and accurate plaque tracking. Picture courtesy of \cite{luo2019deep}.}
\label{fig:ivoct}
\end{figure}

Analysis of atherosclerotic plaque in
clinical application relies on the use of Intravascular Optical Coherence Tomography (IVOCT), in which a continuous and accurate plaque tracking algorithm
is necessary. However, it is challenging to do so due to speckle noise, complex and various intravascular morphology, and a large number of IVOCT images in a pullback. The detected plaque section is represented as a sector with unified radius and the sector is represented
as two-tuples $d = (\Theta_S, \Theta)$, where $\Theta$ denotes the scale (included angle) of the
detected sector, $\Theta_S \in [0, 2 \pi]$ denotes the localization (starting angle on the polar coordinate space) of the detected sector. The eight transform actions are Bidirectional Expansion (BE), Bidirectional Contraction (BC), Contra Rotation (COR), Clockwise Rotation (CLR), Contra Unilateral Expansion (COUE), Clockwise Unilateral Expansion (CLUE), Clockwise Unilateral Contraction (CLUC), and Contra Unilateral Contraction (COUC). The reward function is defined as
\begin{equation}
    R =
    \begin{cases}
    1 & if~IOU(d^a,g)-IOU(d,g) > 0;\\
    -1 & if~IOU(d^a,g)-IOU(d,g)< 0; \\
    1 & if~IOU(d^a,g)-IOU(d,g) = 0~ \\ & \&~ IOU(d^a,g)>0.95;\\
    -1 & if~IOU(d^a,g)-IOU(d,g) = 0~ \\ & \&~ IOU(d^a,g) < 0.95,\\
    \end{cases}
\end{equation}
where $g$ is the ground truth sector region, $d$ is the current detected sector, and $d^a$ is the next detected sector based on current selected action.
$IOU(d^a, g))=IOU(d, g)$ only happens when stop action is selected. Fig.~\ref{fig:ivoct} is the proposed DRL framework.

\subsubsection{Vessel centerline extraction}

\cite{zhang2018deep} propose to use deep reinforcement learning for vessel centerline tracing in multi-modality 3D volumes. The ground truth vessel center points are given as $\bf G=[g_0,g_1,\ldots,g_n]$.
The key idea is to learn a navigation model for an agent to trace the vessel centerline through an optimal trajectory $\bf P=[p_0,p_1,\ldots,p_m]$. The action space is defined as ${\cal A}=\{left, right, top, bottom, front, back\}$, that is, moving to one of six neighboring voxels.

For the current point ${\bf p_t}$, a corresponding point ${\bf g_d}$ on the centerline that has the minimum distance to the point ${\bf p_t}$ is first found. A point-to-curve measure is then defined as
\begin{equation}
    D( {\bf p_t,G})=\|\lambda( {\bf p_t-g_{d+1}}) + (1-\lambda)( {\bf g_{d+2}-g_{d}}) \|.
\end{equation}
It consists of two terms. The first term pulls the agent position towards the ground truth centerline and the second term enforces the agent towards the direction of the curve. With the aid of $D( {\bf p_t,G})$, the reward function is given as
\begin{equation}
    r_t =
    \begin{cases}
    D( {\bf p_t,G})-D( {\bf p_{t+1},G}), & if~\|{\bf p_t-g_d}\| \le l\\
    \|{\bf p_t-g_d}\| - \|{\bf p_{t+1}-g_d}\|, & otherwise
    \end{cases}
\end{equation}

\begin{figure}
\centering
\includegraphics[width=\columnwidth]{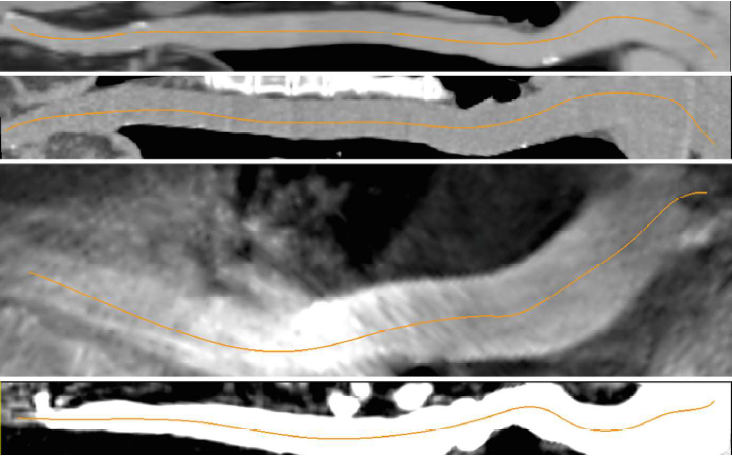}
\caption{Example of traced aorta centerlines in the curved planar reformatting (CPR) view. Picture courtesy of \cite{zhang2018deep}.}
\label{fig:vessel}
\end{figure}

For evaluation, the authors collect 531 contrasted CT, 887 non-contrasted CT, 737 C-arm CT, and 232 MR volumes from multiple sites over the world. For the original 12-bit images, the voxel intensity is clipped and normalized within [500,2000]. The intensity distribution of MR is mapped to that of CT. All these volumes are then mixed for training and testing. The proposed algorithm achieves better performance when compared with a supervised 3D CNN approach.

Recently, \cite{Yuyang_2020_DQN} make use of DDQN and  3D dilated CNN to address the problem of accurate coronary artery centerline. Their network consists of two parts: a DDQN-based tracker to predict the next action and a branch-aware detector to detect the branch points and radius of coronary artery. With such network architecture, it requires only one seed as input to extract an entire coronary tree. The two-branch network has been evaluated on CAT08 challenge and obtains the state-of-the-art performance while it costs only 7s for inference. Fig.~\ref{fig:vessel} shows an example of traced aorta centerlines in the curved planar reformatting (CPR) view.

\subsection{Solving optimization using DRL}

Because DRL can handle the non-differential metrics, it is widely used to solve optimization problems where conventional methods fall apart. Table~\ref{tab:ref2} is an array of such applications including tuning hyperparameters for radiotherapy planning, selecting the right image augmentation selection for image classification, searching best neural architecture for segmentation, and avoiding poor images via a learned acquisition strategy.

\subsubsection{Image classification}

\cite{akrout2019improving} propose to integrate a CNN classification model with a RL-based Question Answering (QA) agent for skin disease classification. To better identify the underlying condition, the DNN-based agent learns how to ask the patient about the presence of symptoms, using the visual information provided by CNN and the answers to the asked questions. It is demonstrated that the integrated approach increases the classification accuracy
over 20\% when compared to the CNN-only approach that uses only
the visual information. It narrows down
the diagnosis faster in terms of the average number of asked questions, when compared with a conventional decision-tree-based QA agent.

\begin{figure}[htbp]
\centering
\includegraphics[width=\columnwidth]{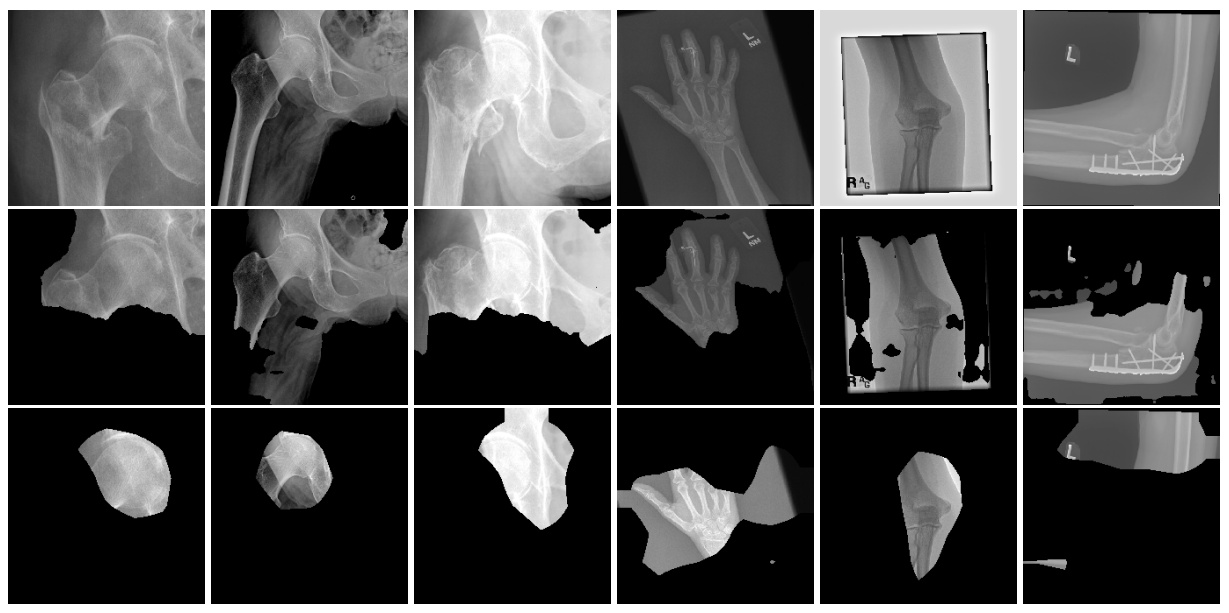}
\caption{X-Ray examples (top) and the masks created by ~\cite{cheng2019adversarial} (middle) and
DenseNet+GradCam (bottom) for hip, hand, and elbow. Picture courtesy of \cite{cheng2019adversarial}.}
\label{fig:xrayhip}
\end{figure}

\cite{cheng2019adversarial} study how to use semantic segmentation that produces a hard attention map for improved classification performance. In particular, a segmentation agent and a classification model are jointly learned. The segmentation agent, which produces a segmentation mask, is trained via a reinforcement learning framework, with reward being the classification accuracy. The classification model is learned using both original and masked data as inputs. Promising results are obtained on Stanford MURA dataset, consisting of 14,863 musculoskeletal studies of elbows, finger, forearm, hand, humerus, shoulder, and wrist with 9,045 normal and
5,818 abnormal labeled cases and on a hip fracture dataset, consisting of 1,118 pelvic radiographs with 6 classes: no fracture, intertrochanteric fracture, displaced femoral neck fracture, non-displaced femoral neck fracture, arthroplasty, and ORIF (previous internal fixation). Fig.~\ref{fig:xrayhip} shows some sample X-Ray images and their corresponding attention maps.

To combat the issue of data shortage in medical image classification, synthesizing realistic medical images offers a viable solution. \cite{ye2020synthetic} investigate the issue of  synthetic sample selection for improved image classification in order to assure the quality of synthetic images for data augmentation purposes because some of the generated images are not realistic and pollute the data distribution. The authors train a DRL agent via proximal policy optimization (PPO) to choose synthetic images containing reliable and informative features, using the classification accuracy as the reward. Extensive experiments are conducted on two image datasets of cervical and lymph node histopathology images and the performances are improved by 8.1\% and 2.3\%, respectively.

\cite{Wang_2020_autoweight} combine four different types of ultrasonography to discriminate between benign and malignant breast nodules by proposing a multi-modal network. In their network, the modalities interact through a RL framework under weight-sharing, i.e., automatically find the optimal weighting across modalities to increase accuracy. Corresponding to four modalities, there are four streams (ResNet18 is used as backbone) and each stream provides one loss. Together four losses from four streams, there is another fusion loss. All the five losses are weighted by coefficients which are automatically learnt through an RL framework. The auto-weighting network is evaluated on 1,616 sets of multi-modal ultrasound images of breast nodules and it shows that multi-modal methods outperform single-modal methods.

\subsubsection{Image segmentation}

Medical image segmentation aims at finding the exact boundary of an anatomical or pathological structure in a medical image. In the most general form an image segmentation approach assigns semantic labels to pixels. By grouping the pixels with the same label, object segmentation is realized. From image segmentation, clinical measurements such as  organ volume can be computed and diseases such as enlarged liver can be diagnosed.

\underline{Using RL in traditional image segmentation.} Image thresholding is a simple method of creating segmentation. All pixels above or below a certain threshold form a segmented object. However, the threshold is nontrivial to obtain. \cite{shokri2003using} propose to use reinforcement learning for determining the optimal threshold.

\cite{sahba2006reinforcement} introduce a reinforcement learning framework for medical image segmentation
The idea is to optimally find the appropriate local threshold and structuring element values to segment the prostate in ultrasound images. Reinforcement learning agent takes an ultrasound image as input and takes some actions (i.e., adjusting different thresholds and structuring element values) to change the the quality of segmentation. Since the number of parameters is limited, the Q-value is learned but without using deep learning. The reinforcement learning agent can use this learned Q-value for similar ultrasound images as well.

\cite{wang2013general} present an online reinforcement learning framework for medical image segmentation. The so-called context specific segmentation is first introduced such that the model not only uses a defined objective function but also incorporates the user's intention and prior knowledge. Based on this, a general reinforcement learning based segmentation framework is proposed in order to take user behaviors into account. It is shown that the proposed framework is able to significantly reduce user interaction, while maintaining both segmentation accuracy and consistency.

However, all the above approaches are still based on a limited number of parameters to derive image segmentation results. This severely limits the segmentation performance. However, using a high number of parameters might make the reinforcement learning intractable even with the aid of deep learning.

\underline{DRL based image segmentation.}
Contemporary medical image segmentation methods are based on machine learning~\cite{zhou2010shape} or fully convolutional deep network structures such as U-Net~\cite{ronneberger2015u}. However, there are a few strategic choices to make in U-Net training, such as tuning the learning rate, data augmentations, data pre-processing, etc. Previous methods are based either on extensive experimentation and grid parameter search or heuristics stemming from specific domain knowledge and expertise; \cite{yang2019searching} present a RL searching approach to optimize the training strategy for 3D medical image segmentation, which boosts the performance of the baseline models.

\begin{figure}
\centering
\includegraphics[width=\columnwidth]{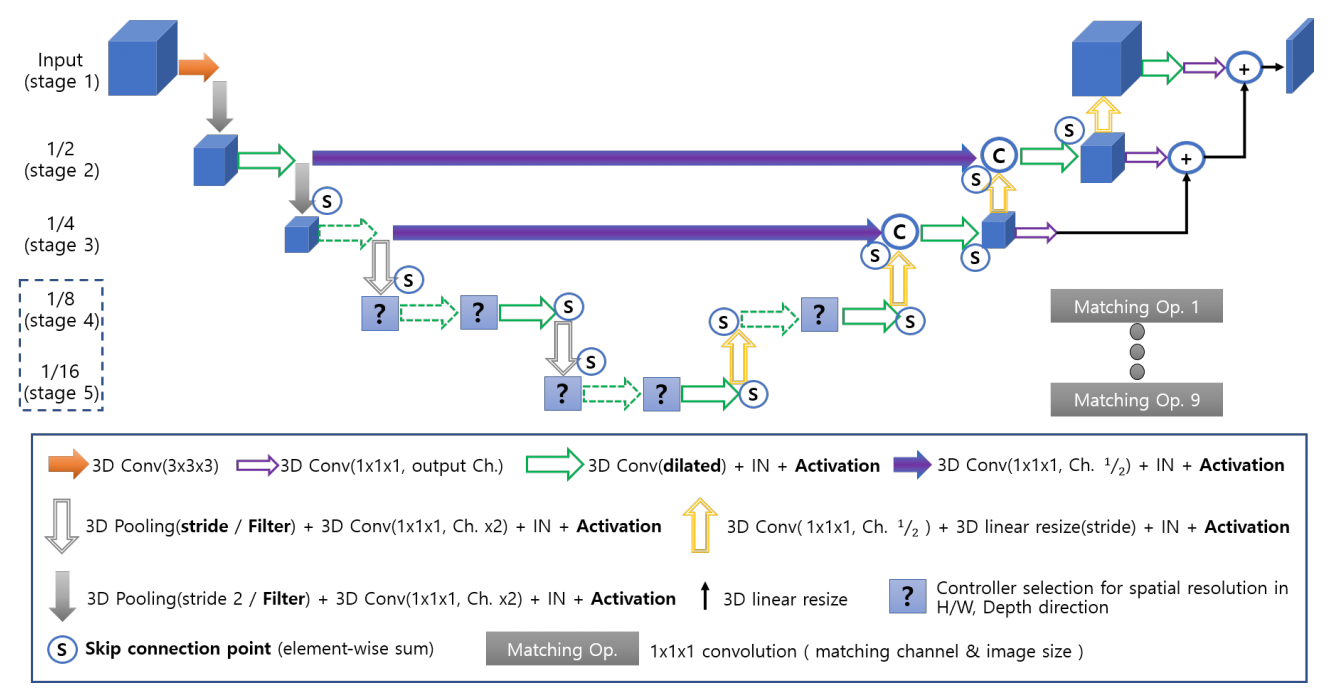}
\caption{The proposed base architecture that is modified to best fit the application by using RL. Picture courtesy of \cite{bae2019resource}.}
\label{F:nas}
\end{figure}

Neural architecture search (NAS)~ \cite{zoph2016neural} automates the task of designing neural networks for a special application, often leading to better performance. However, NAS is seldom applied to medical image segmentation. \cite{bae2019resource} make such an attempt, aiming to modify a U-Net base architecture as in Fig.~\ref{F:nas} so that the image segmentation performance is improved. The search space constitutes multiple factors, including input size, pooling type, filter size, and stride size, activation type, skip connection point, and dilation rate. Using the searched U-Net, the segmentation performances on the medical segmentation decathlon (MSD) challenges are better than those of the nnU-Net approach \cite{isensee2018nnu}, which is considered as the state-of-the-art approach.

The lack of labeled data is one of the most challenges in medical image segmentation. Among existing methods that intend to increase and diversify the available training samples, augmentation has been commonly used~\cite{aug_1, aug_2}. However, data augmentation has been applied as pre-processing and there is no guaranteed that it is optimal. In order to learn an optimal augmentation under an end-to-end segmentation framework, \cite{aug_drl} propose to train both augmentation and segmentation modules simultaneously and use the errors in segmentation procedure as feedback to adjust the augmentation module. In addition to scarce annotation, class-imbalance issue is also addressed in Dual-Unet \cite{dual_unet_2020}, which proposes a semi-supervised approach that leverages RL as a pre-localization step for catheter segmentation. Dual-Unet is trained on both limited labeled and abundant unlabeled images with a two-stage procedure.

Volumetric data is popular in medical analysis while most 3D image segmentation methods usually fail to meet the clinic requirements.  By iteratively incorporating user hints, \cite{multi_agent_iter} propose IteR-MRL with multi-agent reinforcement learning to capture the dependency among voxels for segmentation task as well as to reduce the exploration space to a tractable size.

\subsubsection{Image acquisition and reconstruction}

\begin{figure}[htbp]
\centering
\includegraphics[width=\columnwidth]{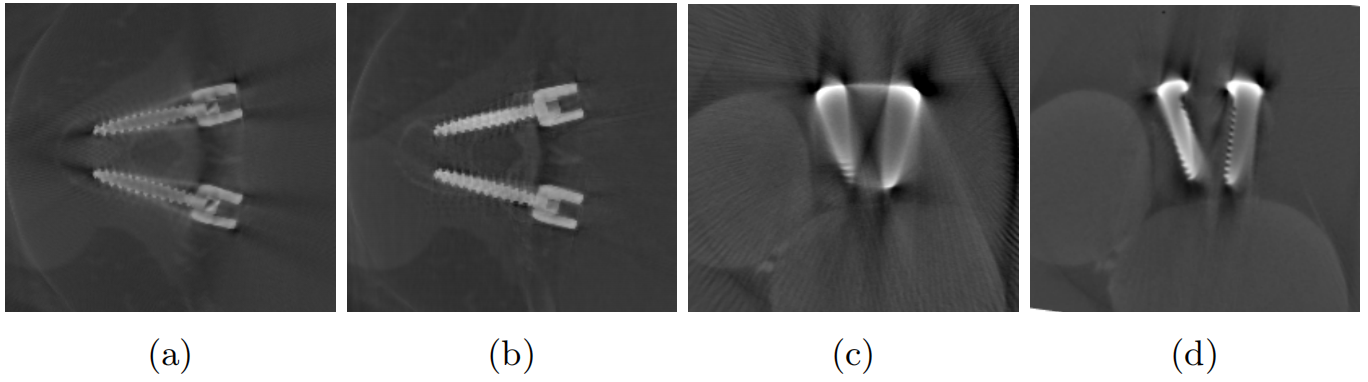}
\caption{Two examples of axial slices from a volume reconstructed from (a,c) a straightforward short-scan and (b,d) a task-aware trajectory recommended by the agent. It is evident that the visual quality of the images reconstructed by using the agent is better. Picture courtesy of \cite{zaech2019learning}.}
\label{fig:metal}
\end{figure}

CT metal artifacts, whose presence affects clinical decision making, are produced because of there is an inconsistency between the imaging physics and idealized assumption used in CT reconstruction algorithm. While there are many metal artifact reduction (MAR) algorithms in the literature that post-process the already acquired data say from a pre-determined cone beam CT imaging trajectory or reconstructed images, \cite{zaech2019learning} propose to design a task-aware, patient-specific imaging trajectory in order to avoid acquiring ``poor'' images that give rise to beam hardening, photon starvation, and noise. Such a design strategy is learned offline via a DRL agent that predicts the next acquisition angle that maximizes a final detectability score. Fig.~\ref{fig:metal} compares the reconstructed images from a straightforward short-scan and a task-aware trajectory recommended by the agent. It is clear that the metal artifacts are reduced.

CT iterative reconstruction solves an optimization problem that say uses a total variation (TV) regularization \cite{rudin1992nonlinear}:
\begin{equation}
    f^* = \arg \min_f \frac{1}{2}|Pf-g|^2 + |\lambda \cdot \nabla f|,
\end{equation}
where $f^*$ is the image to be reconstructed, $P$ is the x-ray projection operator, $g$ is the measured projection signals, $\nabla f$ computes the gradient of the image, and $\lambda$ is a vector of regularization coefficient, which is spatially varying for better modeling. The choice of $\lambda$ is crucial for final image quality; but tuning such parameters is nontrivial. \cite{shen2018intelligent} propose to use a DRL agent that learns a parameter-tuning policy network (PTPN) for such a tuning task. It is demonstrated that, with the aid of the agent, the final image quality reaches a level similar to that with human expert tuning.

\cite{shen2020learning} propose to use DRL to learn a personalized CT scan so that the final reconstructed image quality is maximized, given a fixed dose budget. The key idea is to learn a sequential strategy that selects the acquisition angle and the needed dose for this chosen angle. The reward function is computed as
\begin{equation}
R(s_t, s_{t-}, a_t)
= PSNR(I_t,I)-PSNR(I_{t-1},I), \label{eq:pnsrreward}
\end{equation}
where $I$ is the groundtruth image, $I_t$ is the reconstructed image at time step $t$, and $PSNR(I', I)$ represents
the Peak Signal to Noise Ratio (PSNR) value of the reconstructed image $I'$.
Experiments are conducted using the datasets from 2016 NIH-AAPM-Mayo Clinic Low Dose CT Grand Challenge, demonstrating that the learned scanning policy yields better overall reconstruction results with the acquisition angles and dose are adaptively adjusted.

\cite{pineda2020active} propose to optimize the sequence of k-space measurements, aiming to reduce the number of measurements taken and thus accelerate the acquisition. By formulating it as a partially observable Markov decision process, a policy that maps history of k-space measurements to an index of k-space measurement to acquire next is then learned using DDQN.
Similar to (\ref{eq:pnsrreward}), the reward is defined as the decrease in reconstruction
metric with respect to the previous reconstruction.
Experiments on the fastMRI dataset of knees~\cite{zbontar2018fastmri} demonstrate that the learned policy outperforms other competing policies in terms of final reconstruction quality, over a large range of acceleration factors. Recently, \cite{mri_res} extend pixelRL \cite{pixelrl} by assigning each pixel of the input image an agent that changes the pixel value. In their work, both reinforcement learning techniques and classical image filters are taken into to reconstruct MRI.

\subsubsection{Radiotherapy planning}

Radiotherapy planning often involves optimizing an objective function with constraints, which consists of multiple terms that are weighted.
Weigh adjusting requires expertise from a human expert in order to yield a high quality plan. \cite{shen2018intelligent} leverage DRL to learn a weight-tuning policy network (WTPN) that takes the current dose volume histogram of a plan as input and outputs an action that adjusts weights, with a reward function that promotes the sparing of organs at risk. The agent is then applied for planning the high-dose-rate brachytherapy for five patients, yielding the quality score 10.7\% higher than human planners.

\subsection{Miscellaneous topics}

The below topics are not about analyzing clinical medical images, but they are related in general. What is common among them is that they all use reinforcement learning as a base technology.

\subsubsection{Video summarization}
Recently, \cite{Liu_2020_video} introduce a fully automatic video summarization method using DRL. Their network contains an encoder-decoder CNN to first extract visual representation and then feed the feature into a Bi-LSTM to model time dependency. Finally, the RL network interprets the summarization task as a decision making process and takes actions on whether a frame should be selected for the summary set or not. In their framework, the reward is defined as the quality of the selected frames in terms of their representation, diversity, as well as the likelihood of being a standard diagnostic view plane. The proposed network can be implemented as either supervised or un-supervised manner and it obtains state-of-the-art summarization performance with highest $F_1$ score

\subsubsection{Surgical gesture segmentation and classification}

In a different kind of application related to medical surgery, DRL is applied to recognize surgical gestures from a video \cite{liu2018deep}. This is different from prior work that is based on graphical models such as HMM and CRF or deep learning models such as recurrent neural network and temporal convolutional network (TCN). \cite{liu2018deep} set up a sequential decision-making problem and solve it using DRL that is built upon the TCN features.

An interesting design is to use different time steps when walking through the video sequence until reaching the end. A small time step $k_s$ is useful when the classification is not discriminative enough such as at the gesture boundaries and a large time step $k_l$ is useful otherwise. Experimental results on the benchmark JIGSAWS dataset demonstrate that the proposed DRL achieves similar performance to TCN. The use of a large time step contributes a higher edit score.

\subsubsection{Personalized mobile health intervention}

The prevalence of smartphones and wearable devices makes
mobile health technology an important research direction which holds promise in impacting people's health. One idea is to use smart devices to collect and analyze raw data and to provide the device users in-time interventions, such as reduced alcohol abuse and obesity management.

Since reinforcement learning offers a sequential decision making framework, it is a natural choice for mobile data analysis. However, such an analysis often assumes that all users share the same RL model or each user has own RL model. \cite{zhu2018group} propose group-driven RL that deals with a more realistic situation: a user may be similar to some, but not all. The core idea is to find the so-called similarity network for users and cluster the users into different groups, with each group learning an RL model.

\subsubsection{Computational model personalization}

Computational multi-physics and multi-scale modeling \cite{krishnamurthy2013patient} can improve patient stratification and therapy planning. However, personalization of such model, that is, the process of fitting a multi-physics computational model to clinical measurements or patient data, is a challenging research problem due to the high complexity of the models and the often noisy and sparse clinical data.

\cite{neumann2015vito,neumann2016self} propose to use an artificial agent for model personalization. Specifically, the agent
learns a decision process model through exploration of the computational model offline, how the model behaves under change of parameters, and an optimal strategy for on-line personalization. In experiments of applying the agent to the inverse problems of cardiac electrophysiology and the personalization of a whole-body circulation model, the proposed algorithm is able to obtain equivalent results to standard methods, while being more robust (up to 11\% higher success rates)
and faster (up to seven times).

Finally, \cite{abdi2018muscle} propose to use reinforcement learning for muscle excitation estimation in biomechanical simulation. \cite{joos2020reinforcement} conduct reinforcement learning for musculoskeletal control from functional simulations.

\section{Conclusions and Future Perspectives}
\label{sec:future}

DRL is a powerful framework for medical image analysis tasks. It has been successfully applied to various tasks, including image-based parameter inference in landmark localization, object detection, and registration. DRL has also been demonstrated to be an effective alternative for solving difficult optimization problems, including tuning parameters, selecting augmentation strategies, and neural architecture search. However, realizing the full potential of DRL for medical imaging requires solving several challenges ahead of us and relying on the adoptions of latest DRL advances.

\subsection{Challenges ahead}
We foresee that successful application of DRL to medical image analysis needs to address the following challenges.

\begin{itemize}
\item \underline{Defining a reward function.} It is usually hard to define a reward function for the task at hand because it requires the knowledge from different domains that may not always be available. A reward function with too long delay makes training difficult. In contrast, assigning a reward for each action requires careful and manual human design. Furthermore, the intermediate rewards at each time step are not always accessible. Thus, there is no feedback on how to improve the performance during the episode and what action sequences lead to the maximum final reward.

\item \underline{Q-learning when high-dimensional.} Training a Q-function on a high-dimensional and continuous action space is challenging. For this reason, existing works using low-dimensional parameterization, typically less than 10 with an exception \cite{krebs2017robust} that uses 15-D and 25-D to model 2D and 3D registration, respectively. 

\item \underline{Data availability.} DRL requires a large amount of training data or expert demonstrations. Big datasets are expensive and hard to come by, especially in medical domains. Developing more data-efficient DRL algorithms is desirable to make this technology more widely applicable to the medical imaging community. Shifting from supervised to semi-supervised and unsupervised training, as well as from model-free to model-based approaches is promising directions to address the above-mentioned challenges.
 
\item \underline{Dynamic environment.} Currently the approaches we have reviewed assume a stationary environment, from which observations are made. For example, the environment in the landmark detection is the image itself and what is observed is the image patch that is specified by the state (aka the location) and cropped from the image. In such case, the environment is known but an analytic solution is not available, and DRL is used to find such an approximate solution efficiently. However, the reinforcement learning framework naturally accommodates dynamic environment, that is, the environment itself evolves with the state and action. In other words, the only way to collect information about the environment is to interact with it. Once such example is learning to scan or active acquisition \cite{zaech2019learning,zhang2019reducing,shen2020learning,pineda2020active}, which opens the possibility of personalized scan with a even faster speed and at a more reduced dose. However, currently the existing works demonstrate the idea using simulated environment, future works using real data from real scanning scenarios are needed.

\item \underline{User interaction.} Another aspect worth more attention is user interaction. In the context of parametric medical image analysis, the user input essentially is an external force to escape from the local minimum trap, which gives rise to current result. However, the subsequent behavior after escaping is largely unexplored.

\item \underline{Reproducibility.} Reproducibility is another issue. According to \cite{henderson2017deep}, reproducing existing DRL work is not a straightforward task because there are non-deterministic factors even in standard
benchmark environments and intrinsic variations with respect to specific methods. This statement also holds for DRL in medical imaging.

\end{itemize}

\subsection{The latest DRL advances}


The following latest DRL advances are worth attention and may promote new insights for many medical image analysis tasks.

\begin{itemize}
    \item \underline{Inverse DRL.} DRL has been successfully applied into domains where the reward function is clearly defined. Defining such a reward function for real-world applications is challenging as it requires the knowledge from different domains that may not always be available. An example is autonomous driving, the reward function should be based on all factors such as driver's behavior, gas consumption, time, speed, safety, driving quality etc. In real world scenario, it is hard to have a control of all these factors. Different from DRL, inverse DRL \cite{inverseRL_2020}, \cite{inverseDRL}, 
    a specific form of imitation learning \cite{imiation_learning}, infers the reward function of an agent, given its policy or observed behavior, thereby avoiding a manual specification of its reward function. In the same problem of autonomous driving, inverse RL first uses a dataset collected from the human-generated driving and then approximates the reward function for the task. Inverse RL has been successfully applied to many domains \cite{inverseDRL}. Recently, to analyze complex human movement and control high-dimensional robot systems, \cite{online_IRL} propose an online inverse RL algorithm. In \cite{RL_IRL}, both RL and inverse RL are combined to address planning problem in autonomous driving. 
    
    \item \underline{Multi-Agent DRL.} Most of the successful DRL applications such as game~\cite{multi-agent-game}, \cite{multi-agent-game2}, robotics~\cite{multi-agent-robotics}, autonomous driving \cite{multi-agent-autonomous}, stock trading \cite{MARL_stock}, and social science~\cite{MARL_social} involve multiple players and  require a model with multiple agents. Take autonomous driving as an instance, multi-agent DRL addresses the sequential decision-making problem which involves many autonomous agents, each of which aims to optimize its own utility return by interacting with the environment and other agents \cite{MARL_autonomous}. Learning in a multi-agent scenario is more difficult than a single-agent scenario because of non-stationarity \cite{hernandez2017survey}, multi-dimensionality \cite{MARL_autonomous}, credit assignment \cite{wolpert2002optimal} etc. Depend on whether the multi-agent DRL approach is either fully cooperative or fully competitive, The agents can either collaborate to optimize a long-term utility or compete so that the utility is summed to zero. Recent work on Multi-Agent RL pays attention on learning a new criteria or new setup \cite{new_multi_RL}. 
    
    \item \underline{Meta RL.} As aforementioned, DRL algorithms consume large amounts of experience in order to learn an individual task and are unable to generalize the learned policy to newer problems. To alleviate the data challenge, Meta-RL algorithms \cite{schweighofer2003meta}, \cite{meta_RL} are studied to enable agents to learn new skills from small amounts of experience. Recently there is a research interest in meta RL~\cite{nagabandi2018learning}, \cite{gupta2018meta}, \cite{saemundsson2018meta}, \cite{rakelly2019efficient}, \cite{liu2019taming}, each using a different approach. For benchmarking and evaluation of meta RL algorithms, \cite{yu2020meta} present Meta-world, which is an open-source simulator consisting of 50 distinct robotic manipulation tasks.
    
    \item \underline{Imitation Learning.} Imitation learning is close to learning from demonstrations which aims at training a policy to mimic an expert's behavior given the samples collected from that expert. Imitation learning is also considered as an alternative to RL/DRL to solve sequential decision-making problems. Beside inverse DRL, an imitation learning approach as aforementioned, behavior cloning is another an imitation learning approach to train policy under supervise learning manner.  \cite{imitation_1} present a method for unsupervised third-person imitation learning to observe how other humans perform tasks. Building on top of Deep Deterministic Policy Gradients and Hindsight Experience Replay, \cite{clone_DRL} propose a behavior cloning loss function to increase the level of imitating the demonstrations. Besides Q-learning, Generative Adversarial Imitation Learning \cite{gail_IL} propose P-GAIL that integrate imitation learning into the policy gradient framework. P-GAIL considers both smoothness of policy update and the diversity of the learned policy by utilizing Deep P-Network \cite{P_Dnn}.
\end{itemize}

\section*{Credit authorship contribution statement }
\textbf{S. Kevin Zhou}: Conceptualization, Writing original draft, Writing - review and editing.
\textbf{Hoang Ngan Le}: Conceptualization, Writing original draft, Writing - review and editing.
\textbf{Khoa Luu}: Conceptualization, Writing original draft, Writing - review and editing.
\textbf{Hien V. Nguyen}: Conceptualization, Writing original draft, Writing - review and editing.
\textbf{Nicholas Ayache}: Writing - review and editing.

\bibliographystyle{model2-names.bst}
\biboptions{authoryear}
\bibliography{main}
\end{document}